\documentclass[aps,prb,10pt,amsmath,amssymb,superscriptaddress,twocolumn,notitlepage,nofootinbib, longbibliography]{revtex4-2}

\PassOptionsToPackage{pdftex,
pdfencoding=auto,
pdfnewwindow=true,
pdfusetitle=true,
bookmarks=true,
bookmarksnumbered=true,
bookmarksopen=true,
pdfpagemode=UseThumbs,
bookmarksopenlevel=1,
pdfpagelabels=true,
breaklinks=true,
colorlinks=true,
}{hyperref}

\usepackage[utf8]{inputenx}
\usepackage[english]{babel}
\usepackage{graphicx}
\usepackage[usenames,dvipsnames]{xcolor}
\definecolor{purple}{RGB}{128,0,128}
\definecolor{ultramarine}{RGB}{63, 0, 255}
\definecolor{medblue}{RGB}{0, 0, 100}
\definecolor{googleblue}{RGB}{34, 0, 204}
\definecolor{panblue}{RGB}{0,24,150}
\definecolor{carmine}{RGB}{150, 0, 24}
\definecolor{gray}{RGB}{150, 150, 150}
\definecolor{darkgreen}{RGB}{0, 80, 0}

\usepackage[]{hyperref}
\hypersetup{linkcolor=carmine,citecolor=darkgreen,urlcolor=googleblue,anchorcolor=OliveGreen}

\usepackage{microtype}
\microtypecontext{spacing=nonfrench}
\microtypesetup{
expansion={true,nocompatibility},
protrusion={true,nocompatibility},
activate={true,nocompatibility},
tracking=true,
kerning=true,
spacing={true}
}
\usepackage[all=normal,floats=tight,mathspacing=tight,wordspacing=tight,paragraphs=normal,tracking=tight,charwidths=tight]{savetrees}
\everypar=\expandafter{\the\everypar\loosness=-1 }
\usepackage[style=american,autopunct=true]{csquotes} 
\usepackage{mathtools}

\usepackage{soul}

\newcommand{\ket}[1]{\left|#1\right\rangle}

\usepackage{subfigure}

\usepackage{verbatim}
\usepackage{verbatimbox}

\usepackage{floatrow}

\usepackage{booktabs}
\usepackage{times,bbm,amsmath,amssymb}
\usepackage{epsfig,color}
\usepackage{xcolor}
\usepackage{hyperref}
\usepackage{float,siunitx}
\usepackage{thumbpdf,enumerate}
\usepackage{booktabs}
\usepackage{sidecap}
\usepackage[scaled=.8]{couriers}
\usepackage{pstricks}
\usepackage{multirow}
\usepackage{placeins}
\usepackage{pst-grad,bm}
\usepackage{epigraph}
\usepackage{textcomp}
\usepackage{gensymb}
\usepackage{gensymb}
\usepackage{soul}
\usepackage{braket}
\usepackage{acronym}
\usepackage{enumitem}
\usepackage{float}

\usepackage{acronym}

\usepackage{physics}

\usepackage{tikz}

\DeclareUnicodeCharacter{2009}{\,}

\usepackage{soul}

\begin{document}
\onecolumngrid
\twocolumngrid

\title{Modular Quantum-to-Quantum Bernoulli Factory in an Integrated Photonic Processor}

\author{Francesco Hoch} 
\affiliation{Dipartimento di Fisica, Sapienza Universit\`{a} di Roma,
Piazzale Aldo Moro 5, I-00185 Roma, Italy}

\author{Taira Giordani} 
\affiliation{Dipartimento di Fisica, Sapienza Universit\`{a} di Roma,
Piazzale Aldo Moro 5, I-00185 Roma, Italy}

\author{Luca Castello} 
\affiliation{Dipartimento di Fisica, Sapienza Universit\`{a} di Roma,
Piazzale Aldo Moro 5, I-00185 Roma, Italy}

\author{Gonzalo Carvacho}
\affiliation{Dipartimento di Fisica, Sapienza Universit\`{a} di Roma,
Piazzale Aldo Moro 5, I-00185 Roma, Italy}

\author{Nicol\`o Spagnolo}
\affiliation{Dipartimento di Fisica, Sapienza Universit\`{a} di Roma,
Piazzale Aldo Moro 5, I-00185 Roma, Italy}

\author{Francesco Ceccarelli}
\affiliation{Istituto di Fotonica e Nanotecnologie, Consiglio Nazionale delle Ricerche (IFN-CNR), 
Piazza Leonardo da Vinci, 32, I-20133 Milano, Italy}

\author{Ciro Pentangelo}
\affiliation{Istituto di Fotonica e Nanotecnologie, Consiglio Nazionale delle Ricerche (IFN-CNR), 
Piazza Leonardo da Vinci, 32, I-20133 Milano, Italy}
\affiliation{Dipartimento di Fisica, Politecnico di Milano, Piazza Leonardo da Vinci, 32, I-20133 Milano, Italy}

\author{Simone Piacentini}
\affiliation{Istituto di Fotonica e Nanotecnologie, Consiglio Nazionale delle Ricerche (IFN-CNR), 
Piazza Leonardo da Vinci, 32, I-20133 Milano, Italy}

\author{Andrea Crespi}
\affiliation{Istituto di Fotonica e Nanotecnologie, Consiglio Nazionale delle Ricerche (IFN-CNR), 
Piazza Leonardo da Vinci, 32, I-20133 Milano, Italy}
\affiliation{Dipartimento di Fisica, Politecnico di Milano, Piazza Leonardo da Vinci, 32, I-20133 Milano, Italy}

\author{Roberto Osellame}
\affiliation{Istituto di Fotonica e Nanotecnologie, Consiglio Nazionale delle Ricerche (IFN-CNR), 
Piazza Leonardo da Vinci, 32, I-20133 Milano, Italy}

\author{Ernesto F. Galv\~ao}
\affiliation{International Iberian Nanotechnology Laboratory (INL)\\
 Av. Mestre José Veiga s/n, 4715-330 Braga, Portugal
}
\affiliation{Instituto de F\'isica, Universidade Federal Fluminense, Av. Gal. Milton Tavares de Souza s/n, Niter\'oi, RJ, 24210-340, Brazil}

\author{Fabio Sciarrino}
\email[Corresponding author: ]{fabio.sciarrino@uniroma1.it}
\affiliation{Dipartimento di Fisica, Sapienza Universit\`{a} di Roma,
Piazzale Aldo Moro 5, I-00185 Roma, Italy}

\begin{abstract}
Generation and manipulation of randomness is a relevant task for several applications of information technology. It has been shown that quantum mechanics offers some advantages for this type of task. A promising model for randomness manipulation is provided by the Bernoulli factories, protocols capable of changing the bias of Bernoulli random processes in a controlled way. At first, this framework was proposed and investigated in a fully classical regime. Recent extensions of this model to the quantum case showed the possibility of implementing a wider class of randomness manipulation functions. We propose a Bernoulli factory scheme with quantum states as input and output, using a photonic path-encoding approach. Our scheme is modular, universal, and its functioning is truly oblivious of the input bias, characteristics that were missing in earlier work. We report on experimental implementations using an integrated and fully programmable photonic platform, thus demonstrating the viability of our approach. These results open new paths for randomness manipulation with integrated quantum technologies.
\end{abstract}

\maketitle

\section{Introduction}

Randomness plays an essential role in several research fields and daily life applications, such as those connected to sensitive data protection. There are several deterministic techniques that can be exploited to generate randomness, whose security and efficiency depend on the precise algorithm used. Quantum mechanics provides intrinsic randomness, which is unbreakable from the theoretical point of view, but hard to ensure from the experimental one, due to the inevitable noise and imperfect control over devices. This peculiar property of quantum theory leads to several advantages in the manipulation, communication and processing of information, which are shown by various quantum communication protocols \cite{PhysRevLett.67.661,BENNETT20147,Harrow2017} and quantum computational algorithms \cite{Feynman1982, Nielsen2009,Shor1997,BENNETT20147,Grover1996,Shor1997,Deutsch1992}. The generation and manipulation of quantum randomness have been studied in depth, resulting in implementations using different platforms \cite{Blok2015,Willett2013,google2019,polino2020}, degrees of freedom \cite{Agresti2020, randomoam,randommult,revrandom} and protocols \cite{v009a004,Boixo2018,Pironio2010,Colbeck_2011,Hamilton2017}.

A recent proposal aims at using quantum resources to manipulate randomness in Bernoulli processes. Classical Bernoulli factories were first introduced by Keane and O’Brien \cite{Keane1994}, to address the problem of how to
process instances of a Bernoulli variable (flips of a biased coin), with the goal of generating an output Bernoulli variable whose bias is a desired function of the (unknown) input bias. This task was called a Classical-to-Classical Bernoulli Factory (CCBF), since both the input and the output are classical coins, and finds applications in several fields ranging from Markov chain Monte Carlo simulation \cite{Vats2021} to economy \cite{Dughmi2017}. In Ref. \cite{Keane1994}, the space of simulable functions was characterized, and a method was proposed to construct them.

In recent years the problem has been extended to the quantum domain by analyzing the possibility of replacing the input and/or the output Bernoulli variables with quantum counterparts. In Ref.~\cite{Dale2015,Dale2016}, the first quantum version of this process, named Quantum-To-Classical Bernoulli Factory (QCBF), was defined by considering a quantum input and a classical output. This QCBF extension simulates a Bernoulli variable given a quantum coin (or quoin) as an input parameter. A quoin is a qubit in a pure state that, when measured in the computational basis, returns a classical Bernoulli variable. It was observed that all functions simulable by a CCBF can also be implemented as a QCBF. Indeed, it is enough to measure the quoin in the computational basis to recover a Bernoulli variable with the same parameter. In Ref. \cite{Dale2015} the authors characterized the space of simulable functions with a quantum input and showed that a change of basis is the only necessary quantum operation required to implement the complete set of simulable functions. In fact, a Bernoulli factory that uses quoins as inputs can implement a strictly larger set of bias manipulation functions than the fully classical case. Moreover, there is experimental evidence that a quantum advantage can be achieved \cite{Yuan2016,Patel2019} with respect to the required number of inputs, even for the class of classically simulable functions. 

A more complex quantum extension of the Bernoulli factory was later proposed by Jiang et al. \cite{Jiang2018}, now having quoins as both input and output and aptly named a Quantum-to-Quantum Bernoulli Factory (QQBF). In Ref. \cite{Jiang2018}, the set of simulable functions by a QQBF was completely characterized, and a procedure to construct them was defined. For any version of Bernoulli factory, it is important that the implementation is the same independently of the input bias, that is, the protocol should not use any information on the bias. Furthermore, any experimental scheme should aim at the possibility of concatenating different operations in a modular fashion without knowledge of the output state from the prior step. All the previous attempts to experimentally implement QQBFs \cite{2002.03076,Zhan2020} were unable to simultaneously enforce these conditions. {Once all the features of the QQBF are verified, the quantum input and output enable its use as a subroutine in quantum algorithms. For example, QQBF-like operations have been used for delegated quantum computing in Ref. \cite{Kashefi2017} to obtain genuine secure quantum state preparation.} 

In this work, we propose a modular approach to implement a genuine QQBF and we report its experimental realization using integrated quantum photonics. In detail, we employ a 6-mode, fully programmable, integrated photonic processor (IPP) to manipulate photonic qubits generated by spontaneous parametric down-conversion (SPDC). Our approach provides a viable route for computational tasks involving Bernoulli processes, within a programmable platform which is highly stable, reliable and compact.

This paper is structured as follows. In Sec.~\ref{sec:Bernoulli} we review the theory of Bernoulli factory processes. Then, in Sec.~\ref{sec:modular_scheme} we describe our proposed modular approach. In Secs.~\ref{sec:implementation_integrated} and \ref{sec:experiment} we discuss our experimental apparatus and we demonstrate both individual and concatenated modules, corresponding to the various operations that lead to a universal QQBF, in principle capable of implementing any quantum simulable function.
 
\section{Bernoulli factory}
\label{sec:Bernoulli}

\begin{figure}[t!]
    \centering
    \includegraphics[width = 0.98\columnwidth]{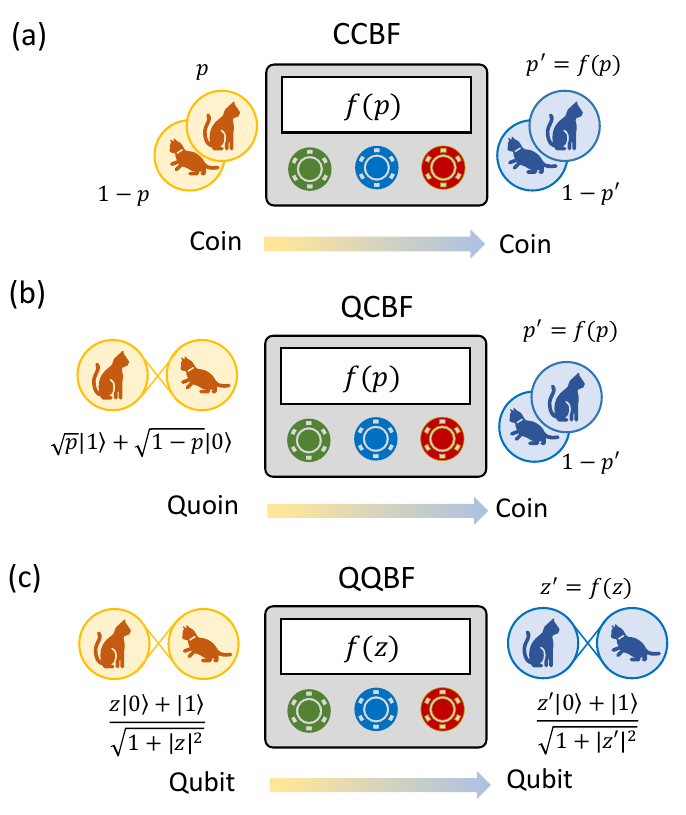}
    \caption{\textbf{Conceptual scheme of a Bernoulli factory.} (a) Classical-to-Classical Bernoulli factory where a sequence of classical coins with unknown bias $p$ are processed for producing a new coin with bias $f(p)$ \cite{Keane1994,Nacu2005}. (b) Quantum-to-Classical version in which a quantum coin serves as input to synthesize a classical one  \cite{Dale2015,Dale2016} and (c) a fully quantum version where both the input and the output are general quantum states \cite{Jiang2018}.}
    \label{fig:Bernoulli}
\end{figure}

Different types of Bernoulli factories are proposed in the literature, which may take either classical or quantum resources as inputs and outputs (see Fig.~\ref{fig:Bernoulli}).

A Bernoulli factory, in a classical context, is an algorithm for the manipulation of random processes that follow a Bernoulli distribution $\mathcal{B}(p)$, described by the bias parameter $p$. More specifically, a Bernoulli factory aims at constructing a function $G_f : \{0, 1\}^{\infty} \rightarrow \{0, 1\}$, associated with a function $f : \mathcal{D} \subseteq [0,1] \rightarrow [0,1]$, such that its application to a sample following a Bernoulli distribution with parameter $p$ is equivalent to sampling exactly from a different Bernoulli distribution with bias parameter $f(p)$. Formally, this corresponds to searching for a function $G_f$ satisfying $G_f(\mathcal{B}(p)^{\infty}) = \mathcal{B}(f(p))$. An essential requirement is that the function $G_f$ must not depend on $p$, which reflects the assumed ignorance of the user about the value of the input bias. In Ref. \cite{Keane1994}, a necessary and sufficient condition for a Bernoulli factory to exist for a given function $f$ was identified. In particular, it was shown that not all functions are exactly implementable as a Bernoulli factory.

The concept of Bernoulli factory has been extended to the quantum domain exploiting a new fundamental resource, namely
a quantum coin, or quoin, of parameter $p$. In detail, a quoin is a qubit in the pure state $\ket{C_p} \coloneqq \sqrt{1-p}\ket{0}+\sqrt{p}\ket{1}$. 
A~QCBF, first proposed in Ref.~\cite{Dale2015}, has quoins as inputs with bias parameter $p$, and produces at the output a series of classical bits that follow a Bernoulli distribution with parameter $f(p)$. The set of functions $f$ for which a CCBF can be constructed was shown \cite{Dale2015,Dale2016} to be strictly included in the set that can be implemented via a QCBF.

On the other hand in the QQBF, introduced in Ref. \cite{Jiang2018}, both input and output are quantum states. In detail, a QQBF takes as input a set of quoins, all with the same bias parameter $p$, and returns a quoin with parameter $f(p):\mathcal{D} \subseteq [0,1] \rightarrow [0,1]$. More in general we define the following parameterization of single-qubit states proved to be helpful in the analysis of Bernoulli factories:
\begin{equation}
\label{eq:qubit}
    \mathbf{\ket{z}} \coloneqq \frac{z\ket{0}+\ket{1}}{\sqrt{1+\abs{z}^2}},
\end{equation}
where $z$ is a complex variable; this can be seen as the stereographic projection of the Bloch sphere onto the complex plane. For a general input qubit $\mathbf{\ket{z}}$ a QQBF associated to a complex function $g(z):\mathbb{C} \rightarrow \mathbb{C}$ is a process that generates at the output a qubit in the state $\mathbf{\ket{g(z)}}$. In Ref. \cite{Jiang2018} it was demonstrated that a necessary and sufficient condition for a QQBF to exist is that the associated function belongs to the complex field generated by the element $z$, i.e. that $g(z)$ is a complex rational function in the parameter $z$. 
{Using the previous result and the algebraic theory of the field, the necessary and sufficient condition to demonstrate the feasibility of implementing all the complex rational functions, i.e. all the simulable QQBF, relies on showing the possibility of implementing the quantum version of the field operations which are inversion, addition and product, and the possibility to combine them.}

\section{Modular scheme for a photonic quantum-to-quantum Bernoulli factory}
\label{sec:modular_scheme}

\begin{figure}[t]
    \centering
    \includegraphics[width=0.99\columnwidth]{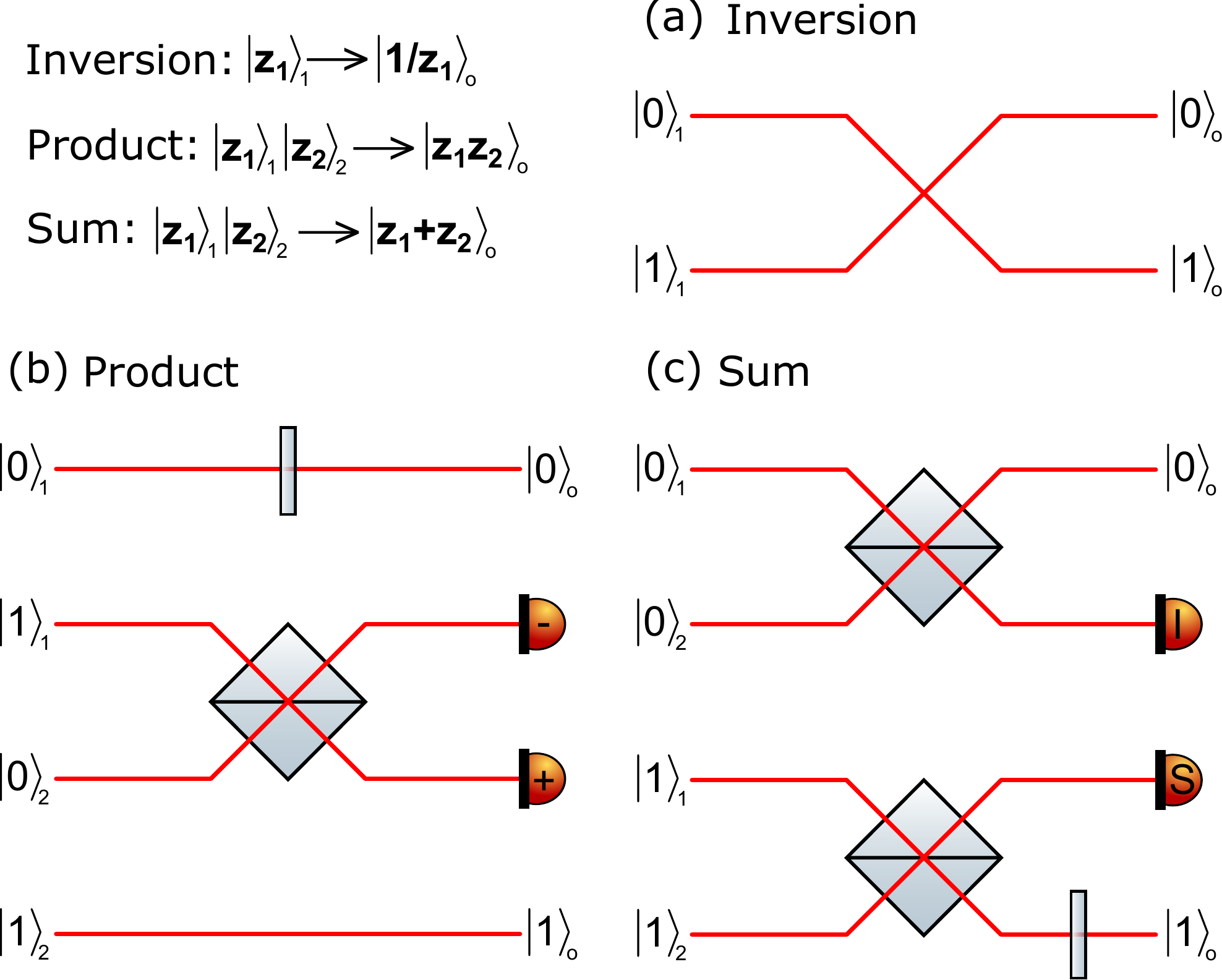}
    \caption{\textbf{Building blocks for a generic Quantum-to-quantum Bernoulli factory.} Interferometric schemes that implement the basic operations to build a generic QQBF with dual-rail encoded qubits. The inputs of the interferometers are labelled by numbers $1$ and $2$ while the outputs are labelled as $O$.
    (a) The inversion operation is performed by swapping the two modes of the input dual-rail qubit. (b) The product operation is performed by sending one waveguide from each dual rail qubit ($\ket{1}_1$ and $\ket{0}_2$) into a balanced BS, and measuring the outgoing modes. Detection of a single photon in the modes labelled "$+$" or "$-$" signals success (up to a global phase). (c) The addition operation is implemented by directing the modes, representing the same state of the two qubits, to equally unbalanced BSs, and measuring one output mode for each BS. When one photon is found in the detector labelled as $S$, and the other photon is in output modes $\vert 0 \rangle_o$ or $\vert 1 \rangle_o$, the output state is the sum of the input ones (up to a global phase). }
    \label{fig:blocchi}
\end{figure}

To demonstrate the feasibility of a generic QQBF using integrated photonics, we will explicitly construct an appropriate scheme to implement the field operations with photons. Previous attempts to experimentally implement the field operations \cite{2002.03076,Zhan2020} were limited, as they substantially relied on prior knowledge of the input state (see Supplementary Note 1). This is in stark contrast to the fundamental requirement for a correct implementation of the protocol, i.e. full ignorance of the input state. Here we present three interferometers (shown in Fig.~\ref{fig:blocchi}), each of them implementing a particular field operation that can be concatenated at will. 
These schemes employ the usual dual-rail encoding for photonic qubits, where logical states $\vert 0 \rangle$ and $\vert 1 \rangle$ are encoded as the presence of a photon in one of two possible optical paths. This choice is motivated by the current state-of-the-art in integrated photonic technology, that allows the implementation of complex architectures \cite{Wang2020} based on beam splitters (BS) and phase shifters.
Let us now discuss the implementation of each field operation building block.

The inversion operation, corresponding to the transformation $\mathbf{\ket{z}} \rightarrow \mathbf{\ket{\frac{1}{z}}}$, is performed by swapping the two modes of the dual-rail qubit [see Fig.~\ref{fig:blocchi}(a)]. It should be noted that this is the only unitary operation among the three, thus having a success probability equal to $1$.

The product operation corresponds to the transformation $\mathbf{\ket{z_1}}\mathbf{\ket{z_2}} \rightarrow \mathbf{\ket{z_1z_2}}$ and can be implemented as shown in Fig.~\ref{fig:blocchi}(b). Two photons are injected in the interferometer, one for each dual-rail qubit mode pair ($\vert 0 \rangle_{1}$, $\vert 1 \rangle_1$) and ($\vert 0 \rangle_{2}$, $\vert 1 \rangle_2$). Then, the modes representing states $\ket{1}_1$ and $\ket{0}_2$ of the two dual-rail qubits are routed as input modes of a balanced BS. The output modes after the BS are then measured by using the two detectors labelled "$+$" and "$-$" in the figure. Conditioned on the detection of a single photon in one of the two outputs of the BS, the output state on the remaining modes $\vert 0 \rangle_{o}$ and $\vert 1 \rangle_o$, after inserting a relative $\pi/2$ phase shift, is found to be $\mathbf{\ket{\pm z_1z_2}_o}=\frac{\ket{1}_o \pm z_1z_2\ket{0}_o}{\sqrt{1+\vert z_1z_2 \vert ^2}}$, where the "$+/-$" sign depends on which detector clicks. Hence, the conditional output is found in the product state, up to a state-independent phase factor of $\pi$. The success probabilities $P_+$ and $P_-$ of the two post-selected outputs are given by:
\begin{equation}
P_+ = P_- = \frac{1+\abs{z_1}^2\abs{z_2}^2}{2(1+\abs{z_1}^2)(1+\abs{z_2}^2)}.
\end{equation}
We observe that the success probability is greater than zero for all inputs, except for the pairs $(z_1 = 0, z_2 = \infty)$ and $(z_1 = \infty, z_2 = 0)$. Indeed, for these pairs, the product operation returns an indeterminate form. In Supplementary Note 2, we provide some further analysis of the behaviour of the success probability.

\begin{figure*}[t]
    \centering
    \includegraphics[width=0.99\textwidth]{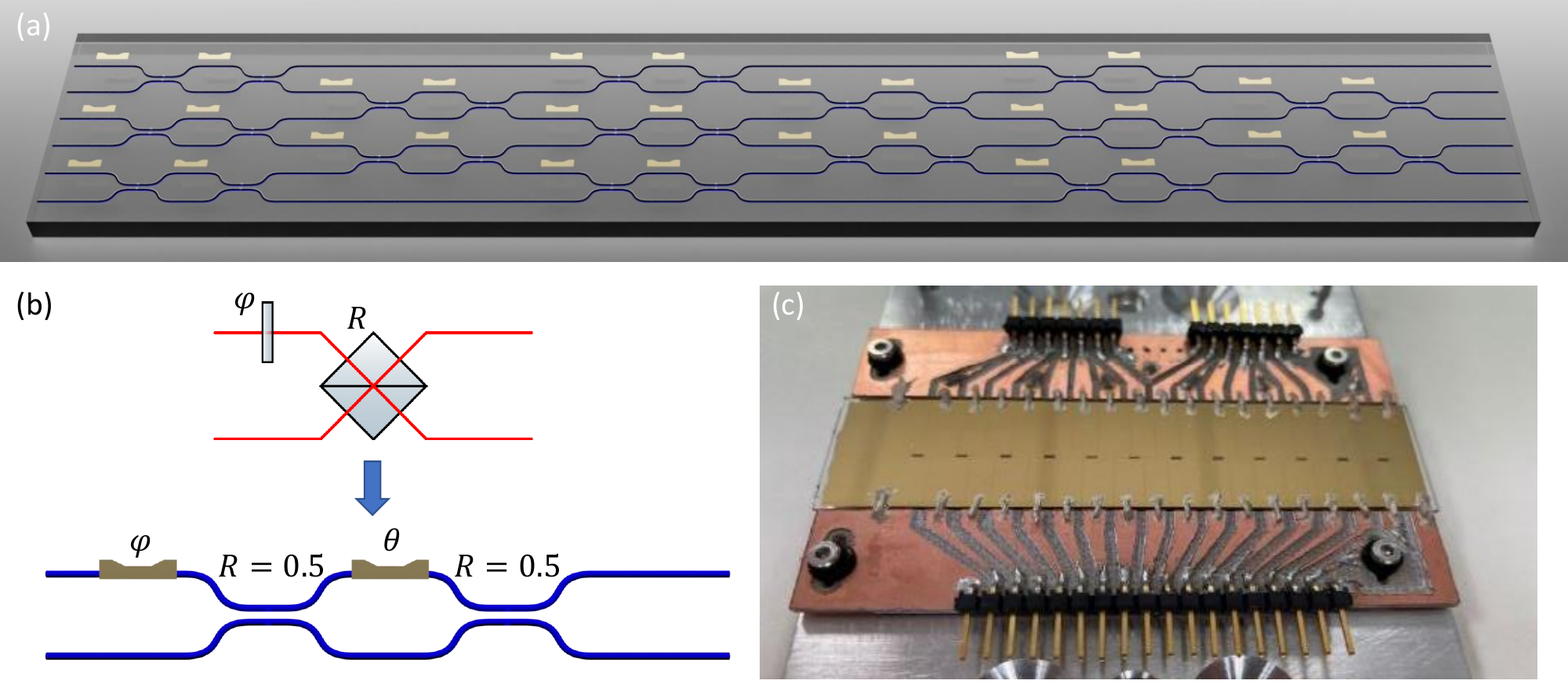}
    \caption{\textbf{Scheme of the 6-mode integrated photonic processor.} (a) Interferometer layout. The device is a 6-mode fully programmable interferometer based on the universal rectangular architecture of Ref. \cite{Clements:16}, allowing the implementation of arbitrary linear-optical transformations. (b) Each beam splitter of arbitrary reflectivity $R$, required in the scheme of Ref. \cite{Clements:16}, is implemented via a module composed of a Mach-Zehnder interferometer with symmetric 50/50 directional couplers, and two tunable phase shifts $\theta$ and $\varphi$. The programmable phases are implemented via thermo-optic phase shifters. (c) Picture of the actual device. {The footprint of the circuit is $82 \times 20\;\text{mm}^2$.} }
    \label{fig:apparato}
\end{figure*}

Finally, the addition operation, corresponding to the transformation $\mathbf{\ket{z_1}}\mathbf{\ket{z_2}} \rightarrow \mathbf{\ket{z_1+z_2}}$, can be implemented with the interferometer of Fig.~\ref{fig:blocchi}(c). Two photons are injected in the interferometer, one for each mode pair ($\vert 0 \rangle_{1}$, $\vert 1 \rangle_1$) and ($\vert 0 \rangle_{2}$, $\vert 1 \rangle_2$). Two identical BSs are used to mix the mode pairs representing the same logical state for the two qubits, combining $\ket{0}_1$ with $\ket{0}_2$ and $\ket{1}_1$ with $\ket{1}_2$. After the mixing process, one output port of each BS is measured via the two detectors labelled as $S$ and $I$ in the figure, while a $\pi/2$ phase shift is added in mode $\ket{1}_o$. Conditioned on the detection of a single photon in $S$, the output state is found to be $\mathbf{\ket{(z_1+z_2)\sqrt{RT}/(R-T)}}_o$, where $R$ and $T$ are the reflectivity and transmissivity of the BSs. If a single photon is detected at $I$ instead, the corresponding output state is $\mathbf{\ket{-z_1 z_2/(z_1+z_2) (R-T)/\sqrt{RT}}}_o$. The numerical multiplicative factor $\sqrt{RT}/(R-T)$ can be set to $1$ by choosing the reflectivity of both BSs to be $R = \frac{5+\sqrt{5}}{10}$. For this choice of $R$, the output conditioned on a click in detector $S$ is the sum state $\mathbf{\ket{z_1+z_2}}_o$, while the one conditioned on a click in detector $I$ is the harmonic mean state $\mathbf{\ket{-z_1 z_2/(z_1+z_2)}}_o$. The corresponding success probabilities are found to be:
\begin{eqnarray}
P_S &=& \frac{\abs{z_1+z_2}^2+1}{5(1+\abs{z_1}^2)(1+\abs{z_2}^2)}, \\ 
P_I &=& \frac{\abs{z_1+z_2}^2+\abs{z_1z_2}^2}{5(1+\abs{z_1}^2)(1+\abs{z_2}^2)}.
\end{eqnarray}
The probability of success is non-zero for all inputs, except for the pairs $(z_1 = \infty, z_2 = \infty)$ for the addition and $(z_1 = 0, z_2 = 0)$ for the harmonic mean, since the results of the corresponding operations for these pairs are an indeterminate form.

{Our implementation, involving linear optics and dual rail-encoding, is thus based on a post-selection process. More specifically, the schemes for the product and addition operations, involving the minimum cost in terms of the number of photons and modes, are found to be probabilistic. Furthermore, the success probability is then found to be dependent on the transformation $f$ being implemented by the scheme, and on the input state $\mathbf{\ket{z_1}}\mathbf{\ket{z_2}}$. However, we observe that such a probabilistic nature is unavoidable within the protocol due to the intrinsic nature of the Bernoulli process.
In Supplementary Note 3 we show that our proposed interferometer designs for the implementation of the field operations are essentially unique if we are to use only four modes.}

After the definition of the building blocks for the presented scheme, we now discuss the possibility of concatenating the field operations. This is an important characteristic feature of our approach, and fundamentally different from previous realizations \cite{2002.03076,Zhan2020}. Our modular scheme allows for a sequential application of the operations. This is possible thanks to the common encoding strategy for the input and the output state of the building blocks. Thus, to concatenate two operations it is sufficient to apply the respective modules in sequence, using the output of one operation as the input of the subsequent one. Importantly, for each product and addition, an additional photon must be added due to the post-selection process required by these operations.

\begin{figure*}[t]
    \centering
    \includegraphics[width=0.99\textwidth]{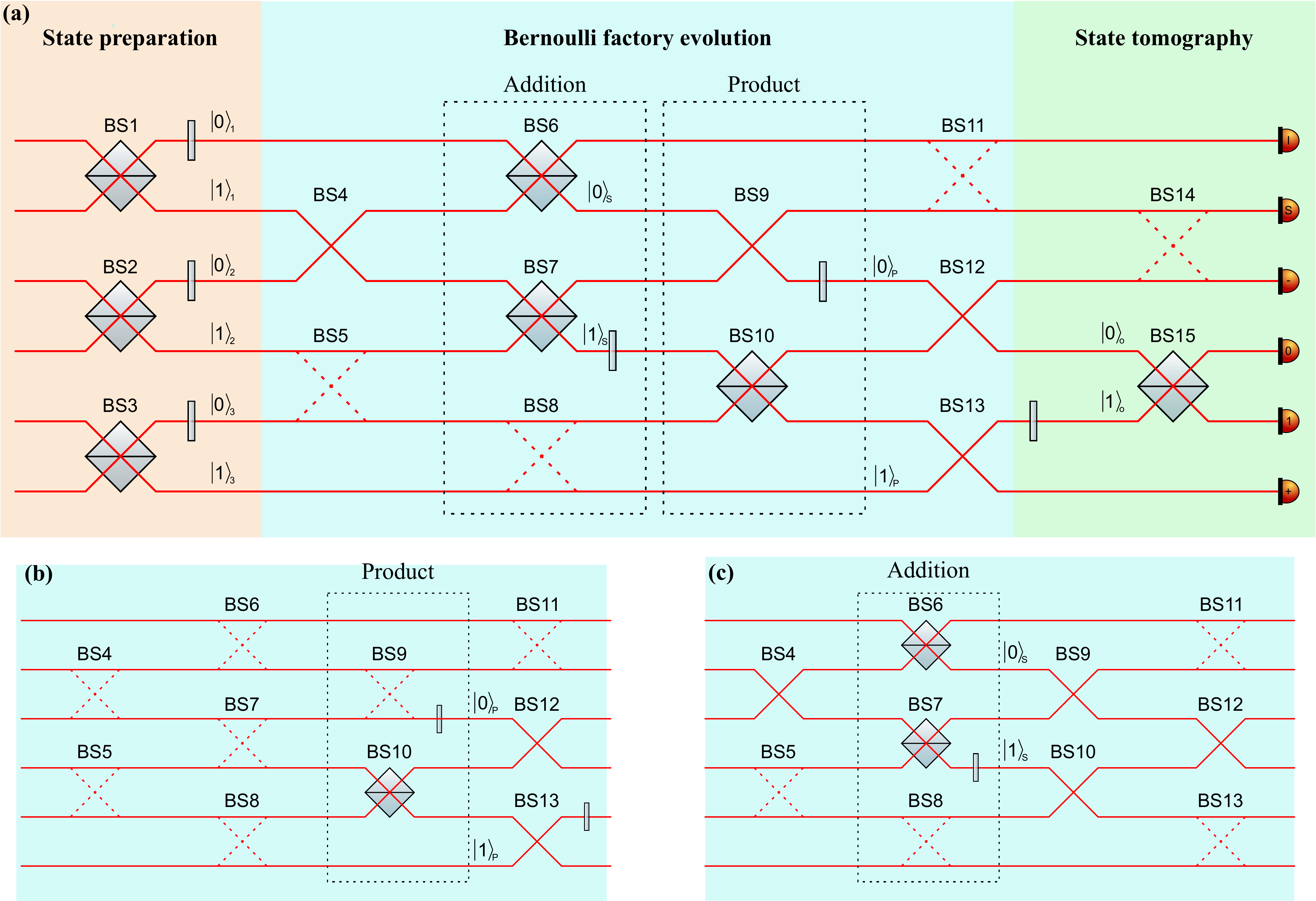}
    \caption{\textbf{Circuit representation of the IPP programmed to implement a complete Bernoulli factory and the building blocks}. (a) Depiction of the full device highlighting stages for state preparation, Bernoulli factory evolution, and state characterization. In particular, the Bernoulli factory evolution shown in this panel corresponds to the settings required to implement the concatenation of an addition followed by a product. In this case, beamsplitters BS5, BS8, BS11 and BS14 are represented with dotted lines since their reflectivities are set to $1$; the reflectivities of BS4, BS9, BS12, BS13 are set to $0$ while the reflectivities of BS6, BS7 and BS10 are tuned to match the value required for the desired operation. BS1, BS2, BS3 and BS15 are controlled during the experiment to generate the input state and reconstruct the output state. (b) Settings of the internal evolution required to implement the building block corresponding to the product operation. (c) Setting corresponding to the required configuration for the addition operation.}
    \label{fig:chip_structure_sum_prod}
\end{figure*}

\section{Implementation}
\label{sec:implementation_integrated}

The experimental certification of our modular QQBF was implemented by using up to 3 photonic qubits in a 6-mode IPP (see Fig. \ref{fig:apparato}). {The IPP was fabricated in-house in a glass substrate by femtosecond laser micromachining \cite{Meany2015, Corrielli2021}.} A complete scheme of the experimental apparatus employed for the experiment is described in the Methods and in Supplementary Note {4}. We discuss now in detail how to implement a QQBF in our 6-mode integrated interferometer, and specifically how to achieve the different required operations by suitably programming the BS network according to the block scheme shown {in} Fig.~\ref{fig:chip_structure_sum_prod}(a). Note that this approach, here demonstrated for 6 modes and thus accommodating for 3 dual-rail qubits, could be extended to arbitrary dimensions by scaling up the architecture. Our 6-mode device is composed of 6 layers of BSs with arbitrary reflectivities, and phases in the $[0, 2\pi)$ interval. The functionality of the different layers can be divided into three main stages, corresponding respectively to: state preparation, Bernoulli factory evolution - implementing the linear optical elements for the desired operation - and state characterization.

In the state preparation stage, the six input modes are mixed in pairs by using three different BSs. For each BS, a phase shifter is present in one of the two output ports. This configuration allows the preparation of a set of generic input qubits in the dual-rail encoding (see Supplementary Note {5}).

In the second stage of the device, the actual evolution for the desired Bernoulli factory operation is applied. More specifically, the reflectivity of the BSs and the phase applied by the phase shifters are appropriately tuned depending on the unitary evolution to be implemented. In particular, the scheme of Fig.~\ref{fig:chip_structure_sum_prod}(a) represents the implementation of an addition operation followed by a product operation, while panels (b) and (c) represent the optical elements required for the implementation of each operation individually (configuration for other operations are reported in the Supplementary Note {6}). Note that the addition operation is similar but not equal to the one represented in Fig.~\ref{fig:blocchi}, since here the first two waveguides are exchanged. This change is inserted to reduce the number of layers required for the concatenation of two operations and is implemented by replacing the reflectivity of BS6 with its complement, thus making BS6 and BS7 complementary.

The final stage performs the necessary operations to characterize the output state. The system can be employed to perform either tomography or direct measurement of the fidelity compared to a target state. State tomography for a single qubit requires three projective measurements on mutually orthogonal bases, from which we can reconstruct the output state \cite{James2001}. On the other hand, to estimate the fidelity the characterization stage is tuned to act as the projector onto the target state {(See Supplementary Note 5). In such a way, verification of the protocol does not require full tomographic reconstruction of the output state.} Note that, if the output state is used as input for additional calculations, the output modes $\vert 0 \rangle_o$ and $\vert 1 \rangle_o$ are not detected and can be routed to subsequent manipulation modules.

\section{Experimental results}
\label{sec:experiment}

A first step towards characterizing the modular QQBF described above involves the demonstration of the individual building blocks by using the 6-mode integrated processor. In particular, according to the required interferometric schemes of Fig. \ref{fig:chip_structure_sum_prod}(b-c), the current inside the thermo-optic phase shifters of the IPP is tuned to provide the required BS reflectivities and phase shifts.

\begin{figure*}[ht!]
    \centering
    \includegraphics[width = \textwidth]{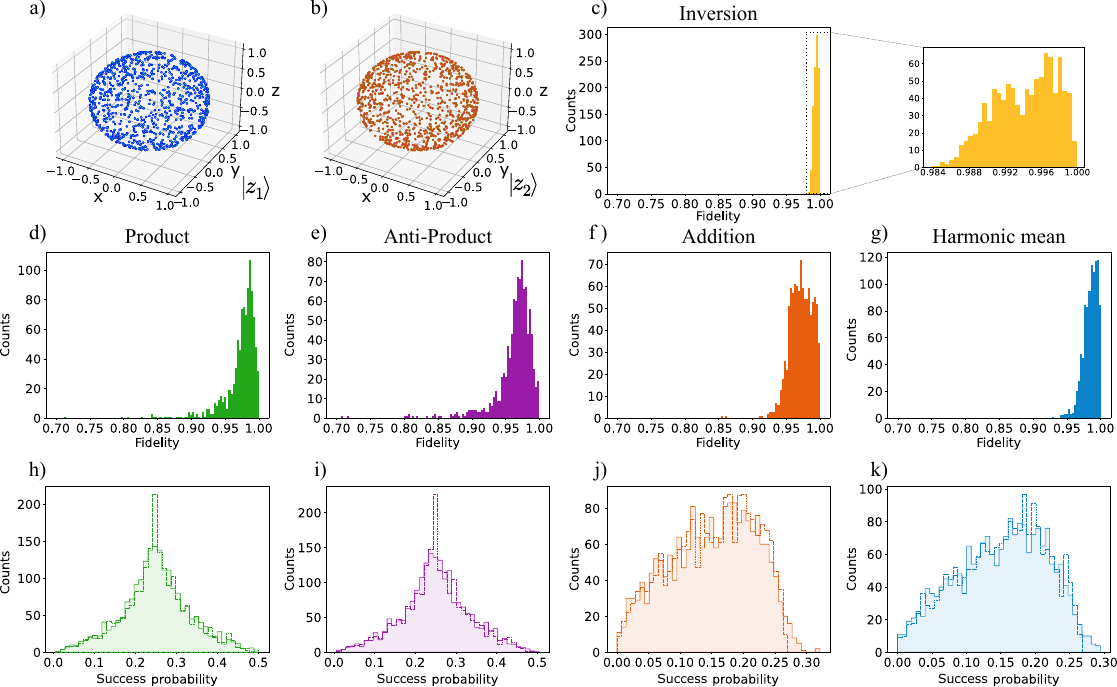}
    \caption{\textbf{Experimental results of the building block operations}. Characterization of the building blocks is performed by generating a set of 1000 pairs of random states $(\vert \mathbf{z_1} \rangle, \vert \mathbf{z_2} \rangle)$ sampled uniformly from the Bloch sphere. (a) Representation of the sampled states $\vert \mathbf{z_1} \rangle$ and in (b) sampled states for $\vert \mathbf{z_2} \rangle$. The fidelity of the output state after the evolution is measured by projecting it onto the known target state. In panels (c-g) we report the distribution of the measured fidelities for each operation, for the set of sampled states. In panels (h-k) we compare the distribution of the success probability for each operation (solid lines) with the corresponding theoretical expectation (dashed lines), for the set of sampled states.}
    \label{fig:Fidelity_single}
\end{figure*}

The operation of every single block is characterized by preparing (in the first stage of the circuit) a set of random input states $(\ket{\mathbf{z_1}},\ket{\mathbf{z_2}})$ sampled from a uniform distribution on the Bloch sphere [see Fig.~\ref{fig:Fidelity_single}(a-b)]. Each pair of states is generated in the state preparation stage of the circuit
by setting the phase and the reflectivity of the first layer of the interferometer.
After the transformation, the output is validated by measuring the success probability of the post-selection used, and the fidelity reached with respect to the target state. The overall figure of merit defining the quality of the implementation is provided by the mean fidelity over the set of sampled states. In Fig. \ref{fig:Fidelity_single}(c-g), we report the results of the measured output fidelities between the output state and the target state of the operation, for all three building blocks (inversion, product, addition). The average results are summarized in Tab.~\ref{tab:concatenation}(a). Furthermore, in Fig. \ref{fig:Fidelity_single}(h-k) we report the histograms showing the output distribution of the success probabilities for the two operations implemented probabilistically (product, addition). From a direct comparison of the obtained results with the theoretical expectations, we find that the operations implemented by the circuit are performed with fidelities close to a unitary value, thus demonstrating the realization of the building blocks of a QQBF. {In this case, corresponding to the verification of each stand-alone operation, the effect of experimental noise due to photon distinguishability is almost negligible. Indeed, the inversion operation scheme does not rely on photon interference, while both product and addition implementations are verified via two-photon experiments, which, in our source, belong to the same generated pair, and thus possess a high degree of indistinguishability.}

As a second step, we demonstrate the modularity of our scheme by showing the possibility of concatenating the individual operations. This aspect is necessary to fulfil all requirements for the correct implementation of a complete Bernoulli factory. To test the concatenation of an addition, followed by a product [$(z_1+z_2)z_3$], the circuit operation is programmed according to the layout of Fig.~\ref{fig:chip_structure_sum_prod}(a). This requires three input photons. The first two photons, impinging respectively in BS1 and BS2, encode the input states for the addition. The third photon impinging on BS3, together with the output state from the first operation, encode the inputs of the product operation. Finally, the output of both concatenated blocks is validated by direct projection onto the target state in the final stage of the device, for an estimation of the fidelity.

To test the correctness of the concatenation, we measure the output fidelity for a particular set of states corresponding to relevant choices of the input. All the results are summarised in Tab.~\ref{tab:concatenation}(b), where we report the obtained output state fidelities. Being a three-photon experiment, this implementation requires the injection of photons generated by the source from different pairs. To compare the experimental data with the theoretical prediction, partial photon distinguishability between the input photons has to be taken into account. Thus, the fidelity $F_C$, after subtraction of dark counts and accidental coincidences, has to be compared with a theoretical model that calculates $F_D$ by taking into account {only} the partial distinguishability between the input photons (see Supplementary Note {7}). We note that this effect is due to the employed photon source, and not to the QQBF implementation itself. The obtained results show a high degree of compatibility between $F_C$ and $F_D$, thus demonstrating the correct implementation of the concatenation of building blocks. Additionally, we have performed the QQBF implementation of a different function, obtained by exchanging the order of the addition and product operations (see Supplementary Note {6}). Also for this different configuration, we obtain a high degree of compatibility with the theoretical predictions. All these results are summarised in Tab.~\ref{tab:concatenation}(c).

\begin{table}[ht!]
\centering
\subtable[Single operations]{
\begin{tabular}{lccc}
    \toprule
        \multirow{2}{*}{\textbf{Operation}} & \multirow{2}{*}{\textbf{Operation}} &\textbf{Measured} & \textbf{Corrected}\\
         & &\textbf{mean Fidelity} & \textbf{mean Fidelity}\\
        & &$F_M$ & $F_C$ \\
    \midrule
        Inversion & $1/z$ & $0.989 \pm 0.003$& $1.000\pm0.003$\\
        Product & $z_1z_2$ &$0.95 \pm 0.02$ & $0.99 \pm 0.02$ \\
        Anti-Product & $-z_1z_2$ & $0.95 \pm 0.03$ & $0.99\pm 0.02$\\
        Addition & $z_1+z_2$ & $0.90 \pm 0.05$ & $0.99 \pm 0.02$ \\
        Harmonic mean & $z_1z_2/(z_1+z_2)$ &$0.92 \pm 0.05$ & $0.99 \pm 0.02$\\
    \bottomrule\\
    \end{tabular}
}

\subtable[Concatenation product-addition]{
    \begin{tabular}{cccc}
    \toprule
        \textbf{Input} & \textbf{Output} & \textbf{Corrected Fidelity} & \textbf{Theoretical Fidelity}\\
        $(z_1,z_2,z_3)$ & $z_1z_2+z_3$ & $F_C$ & $F_D$\\
    \midrule
        $(0,0,0)$ & $0$ & $0.993 \pm 0.005$ & $ 1$\\
        $(\infty,\infty,0)$ & $\infty$ & $0.993 \pm 0.005$ & $ 1$\\
        $(1,1,0)$ & $1$ & $0.95 \pm 0.02$ & $0.96 \pm 0.01$\\
        $(0,0,1)$ & $1$ & $0.80 \pm 0.02$ & $0.79 \pm 0.05$\\
        $(1,1,1)$ & $2$ & $0.93 \pm 0.01$ & $0.92 \pm 0.03$\\
    \bottomrule\\
    \end{tabular}
}
\subtable[Concatenation addition-product]{
    \begin{tabular}{cccc}
    \toprule
        \textbf{Input} & \textbf{Output} & \textbf{Corrected Fidelity} & \textbf{Theoretical Fidelity}\\
        $(z_1,z_2,z_3)$ & $(z_1+z_2)z_3$ & $F_C$ & $F_D$\\
    \midrule
        $(0,0,1)$ & $0$ & $0.98 \pm 0.01$ & $ 1$\\
        $(\infty,0,1)$ & $\infty$ & $1 \pm 0.01$ & $ 1$\\
        $(1,0,1)$ & $1$ & $0.87 \pm 0.02$ & $ 0.88 \pm 0.02$\\
        $(0,1,1)$ & $1$ & $0.88 \pm 0.02$ & $ 0.88 \pm 0.02$\\
        $(1,1,1)$ & $2$ & $0.85 \pm 0.02$ & $ 0.88 \pm 0.02$\\
    \bottomrule\\
    \end{tabular}
}

\caption{\textbf{Summary of the characterization of the building block operations and their concatenation.} (a) The mean fidelity is estimated by averaging over an ensemble of 1000 input sets of states sampled uniformly from the Bloch sphere. (b) Results obtained from the concatenation of a product followed by an addition. (c) Results for the concatenation of an addition, followed by a product. In all the tables, the measured mean fidelity $F_M$ is estimated directly from the raw experimental data. The corrected fidelity $F_C$ is the fidelity measured at the output of the IPP by subtracting dark counts and accidental coincidences. Conversely, $F_D$ is the theoretical fidelity calculated by taking into account the partial distinguishability of the input photons. More details on the data analysis can be found in the Supplementary Notes 7 and {8}. {All the errors are estimated from the propagation of the Poisson statistic proper of single-photon counts.}}
    \label{tab:concatenation}
\end{table}

\section{Discussion}

In this work, we have devised and demonstrated experimentally a full Bernoulli factory working with quantum states both at the input and at the output (i.e. a QQBF). In particular, we have proposed three interferometer designs implementing the basic operations of a field on qubit states. These act as building blocks for the implementation of the Bernoulli factory and, remarkably, can also be concatenated in different orders. This shows the modularity of our approach, making it capable in principle of implementing the complete set of functions known to be theoretically simulable. 
In addition, our methodology guarantees an important ingredient at the core of the Bernoulli factory problem, i.e. manipulations that are truly oblivious to the input state biases. Here
we have implemented our scheme by means of a fully-programmable 6-mode IPP, manipulating three photonic qubits. We report a high degree of control in the optical operation of the IPP and a very high fidelity in the obtained results.

We note that the same device settings allow the implementation of more than one function depending on the post-selection event detected. Further investigation can be foreseen to investigate which functions can be implemented simultaneously with our devices.
{Moreover, the exploitation of fast reconfiguration would enable the application of feed-forward techniques, to allow the programming of subsequent stages depending on measurements and detections performed in previous ones. Feed-forward could enable the active control of phase in the product module, thus, converting the anti-product operation into the product one, and enhancing by a factor 2 the success probability of the operation. Additionally, further investigation involves verifying whether the success probability can be boosted by adding ancillary photons and modes.}

{The successful integration of the algorithm within novel integrated devices, reported in the present manuscript, opens the way to the implementations of the QQBF as a subroutine algorithms in compact platforms which will exploit the stability of the overall process. Indeed}, current developments in photonic integrated technologies are already allowing the realization of systems with progressively increasing sizes. The use of photonic platforms to build a fundamental subroutine allows its natural integration {at the interface between quantum computation and} quantum communication network{s},  {thus enabling the possibility} to exploit the substrate of photonic communication technology, which is presently at a high level of technological and commercial maturity. {In this scenario, the femtosecond laser micromachining technology used to fabricate the IPP may play a significant role in providing custom-tailored photonic components; in particular, its unique 3D capabilities may be beneficial in compactifying circuit designs \cite{meany2016engineering, crespi2016suppression} and in enabling random transformations \cite{hoch2021boson, tang2022generating}.} 
{Moreover, the compatibility of our IPP with different types of photon sources such as demultiplexed quantum dot sources \cite{rodari2024semidevice} allows the possibility to scale up the employed number of photons in coincidence.} {In addition, we can also implement protocols of error mitigation to deal with the experimental imperfection present in the apparatus such as the partial distinguishability between the photons \cite{Marshall2022} or imperfect beam-splitters \cite{Miller2015}.}
The reliability, modularity, and accuracy of our platform pave indeed the way toward the implementation of more complex protocols where Bernoulli factories represent a key ingredient. {Nevertheless, it is worth noting that Bernoulli factories are not limited to photonic implementations, and hence this class of protocols could find application in different platforms ranging from ions to superconducting qubits.}

\section*{Methods}

\textbf{Photon source.} The photons required for the dual-rail encoding are produced by a source based on non-collinear SPDC. In particular, two pairs of photons are emitted by the source, which are deterministically separated in four different paths by exploiting their polarization state (using half-wave plates and polarizing BSs and coupled into single-mode fibers. One photon is  directly detected by a single-photon avalanche photodiode and acts as a trigger. The other three photons are sent through different paths, where they are made indistinguishable in the polarization and time-of-arrival degrees of freedom, and finally injected into the IPP.  Optical fibers are also used to collect light at the outputs of the IPP. The detection stage is composed of 6 in-fiber BSs, one for each output of the IPP, which feeds 12 single-photon avalanche photodiode detectors. In fact, this system implements 6 probabilistic photon-number resolving detectors that are used to characterize the output states.

\textbf{Integrated Photonic Processor}. 
The IPP consists in a reconfigurable, 6-mode waveguide interferometer \cite{pentangelo2024high} realized according to the rectangular layout proposed in Ref. \cite{Clements:16} and thus able to produce any linear transformation of 6 modes. The waveguides follow the optical paths depicted in Fig.~\ref{fig:apparato}(a), the 15 BSs being actually implemented by tunable Mach-Zehnder interferometers {as depicted in Fig.~\ref{fig:apparato}(b)}. Each of the latter components in turn consists of two cascaded waveguide directional couplers and is equipped with two programmable phase shifters. One of the phase shifters is placed inside the Mach-Zehnder ring, while the other one is placed on one of its input ports. By acting on these overall 30 phase shifters full control of the IPP operation is achieved.

We have fabricated the IPP by femtosecond laser micromachining \cite{Meany2015, Corrielli2021} in  EagleXG (Corning Inc.) glass substrate.
The waveguides are directly inscribed in the substrate by laser irradiation, 25~$\mu$m depth below the substrate surface, followed by thermal annealing \cite{Arriola2013, Corrielli2018}. Phase shifters base their functioning on the thermo-optic effect and consist of micro-heaters realized on the chip surface \cite{Flamini2015}. The micro-heaters are resistive paths patterned by laser ablation on a metallic layer, which is deposited on the chip surface. Upon driving suitable currents into the micro-heaters, local heating of the substrate is achieved in a precise and controlled way. Such local heating induces in turn a refractive index change and thus controlled phase delays in the waveguides due to the thermo-optic effect. The chip surface is further micro-structured by femtosecond laser processing, in particular creating thermal-insulation trenches at the sides of the micro-heaters, to increase their efficiency and reduce cross-talks \cite{Ceccarelli2020}. The full IPP has a footprint of $82 \times 20\;\text{mm}^2$. Fiber arrays are glued to the input and output ports to provide optical connections, and fiber-to-fiber optical loss is 3~dB. A careful calibration of the IPP operation with respect to the driving currents in the micro-heaters was performed using coherent light, yielding a fidelity to the target operation on average equal to 99.7\%,
calculated on thousands of randomly chosen unitary transformations. The calibration procedure allowed us to characterize independently the effect of each phase shifter on all the Mach-Zehnder interferometers. The measurements showed a full reconfiguration (i.e. a $2\pi$ phase shift) with a dissipated power of 40.7 $\pm$ 5.4 mW per thermal shifter and a crosstalk on first-neighbour interferometers of 19.0 $\pm$ 5.2\%. More details about the characterization of the processor with classical light can be found in Ref. \cite{pentangelo2024high}.

\section*{Acknowledgements}

This work was supported by the FET project PHOQUSING (“PHOtonic Quantum SamplING machine” - Grant Agreement No. 899544) (F.H., T.G., G.C., N.S., F.C., C.P., S.P., A.C., R.O. E.G., F.S) and by ICSC – Centro Nazionale di Ricerca in High Performance Computing, Big Data and Quantum Computing, funded by European Union – NextGenerationEU (N.S., F.S.). EFG acknowledges support from FCT – Fundação para a Ciência e a Tecnologia (Portugal) via project CEECINST/00062/2018 (E.G.). The IPP was partially fabricated at PoliFAB, the micro- and nanofabrication facility of Politecnico di Milano (\href{https://www.polifab.polimi.it/}{https://www.polifab.polimi.it/}). C.P., F.C. and R.O. wish to thank the PoliFAB staff for their valuable technical support.

\section*{Competing interests}
F.H., T.G., L.C., G.C., N.S., R.O., E.G. and F.S. are listed as inventors on corresponding pending patent applications in Italy (No. 102023000012279) and Portugal (No. 20232005054430) both filed on 15th June 2023 and titled ‘Quantum Bernoulli Factory photonic circuit independent of input state bias’ dealing with schemes for the implementation of Quantum-to-Quantum Bernoulli factory. F.C. and R.O. are co-founders of the company Ephos. The other authors declare no competing interests.
 
\section*{Author contributions}
F.H., T.G., L.C., G.C., N.S., R.O., E.G. and F.S. conceived the concept and the scheme for Quantum-to-Quantum Bernoulli Factory. F.C., C.P., S.P., A.C. and R.O. fabricated the photonic chip and characterized the integrated device using classical optics.  F.H., T.G., G.C., N.S. and F.S. carried out the quantum experiments and performed the data analysis. All the authors discussed the results and contributed to the writing of the paper

\section*{Data availability}
The raw data that support the findings of this study are available from the corresponding authors upon reasonable request.


\end{document}


\tableofcontents

\newpage

\section{Previous experimental implementations}
We now discuss previous schemes and implementations of Quantum-to-Quantum Bernoulli Factories (QQBFs). In the work by Liu et al. \cite{2002.03076}, the authors exploited entangled photons for the building blocks. Each photon generated by the entangled source feeds a displaced Sagnac interferometer and the input state is constructed with a series of waveplates. An important requirement is to have an implementation that is oblivious about the biases of the input states. This assumption is not satisfied in the scheme of Ref. \cite{2002.03076}, since their implementation requires prior knowledge of the state entering the Bernoulli factory process. Indeed this knowledge is exploited to appropriately program the waveplates. Moreover, the same issue is present when trying to concatenate the operations. Indeed, their scheme does not specify how to transfer the quantum information from the output of one operation to the input of the next one in the absence of knowledge about the output state.

A second scheme was proposed by Zhan et al. \cite{Zhan2020}. The authors encoded the two qubits at the input of the building blocks in the same photon: one qubit in the polarization degree of freedom, with the second qubit using dual-rail encoding. The building block then provides an output qubit that is encoded in the polarization degree of freedom of the outgoing photon. The demand for a modular approach that allows for concatenation requires that the output state can be fed as the input of the next operation, along with a new qubit input. In Ref. \cite{Zhan2020}, the authors suggested the following procedure. The output qubit in the polarization degree of freedom is deterministically transferred to a dual-rail encoding, via a birefringent calcite crystal and a half-waveplate. The second qubit is then encoded in the polarization degree of freedom of the same particle via an appropriate set of optical waveplates. Such strategy thus involves setting the input of the second concatenated operation within the Bernoulli factory machine. In order to set the correct configuration of the waveplate, this procedure requires knowledge of the output state of the previous operation, thus not fulfilling the complete set of desiderata for a modular quantum Bernoulli factory.

This concatenation issue in the scheme of Ref. \cite{Zhan2020} can be circumvented by relying on the joining scheme of Ref. \cite{Vitelli2013}. In this  paper, the authors proposed a method to transfer two qubits of information from two distinct particles to two different degrees of freedom of the same one. More specifically, in the implementation of Ref. \cite{Vitelli2013} the two input photons are encoded in the polarization, while the output photon carries the quantum information in the polarization and in a dual-rail qubit encoding. Given that the joining scheme is independent from the input states, it can be used as the intermediate block between two concatenated operations of Ref. \cite{Zhan2020}. However, current linear optical implementations \cite{Vitelli2013} of the joining scheme are probabilistic, and require ancillary photons. Thus, this would correspond to a further reduction of the success rate, and additional overheads, for the QQBF scheme of Ref. \cite{Zhan2020}.

\section{Success probability behavior around the critical points}
\begin{figure}
    \centering
    \includegraphics[width = \textwidth]{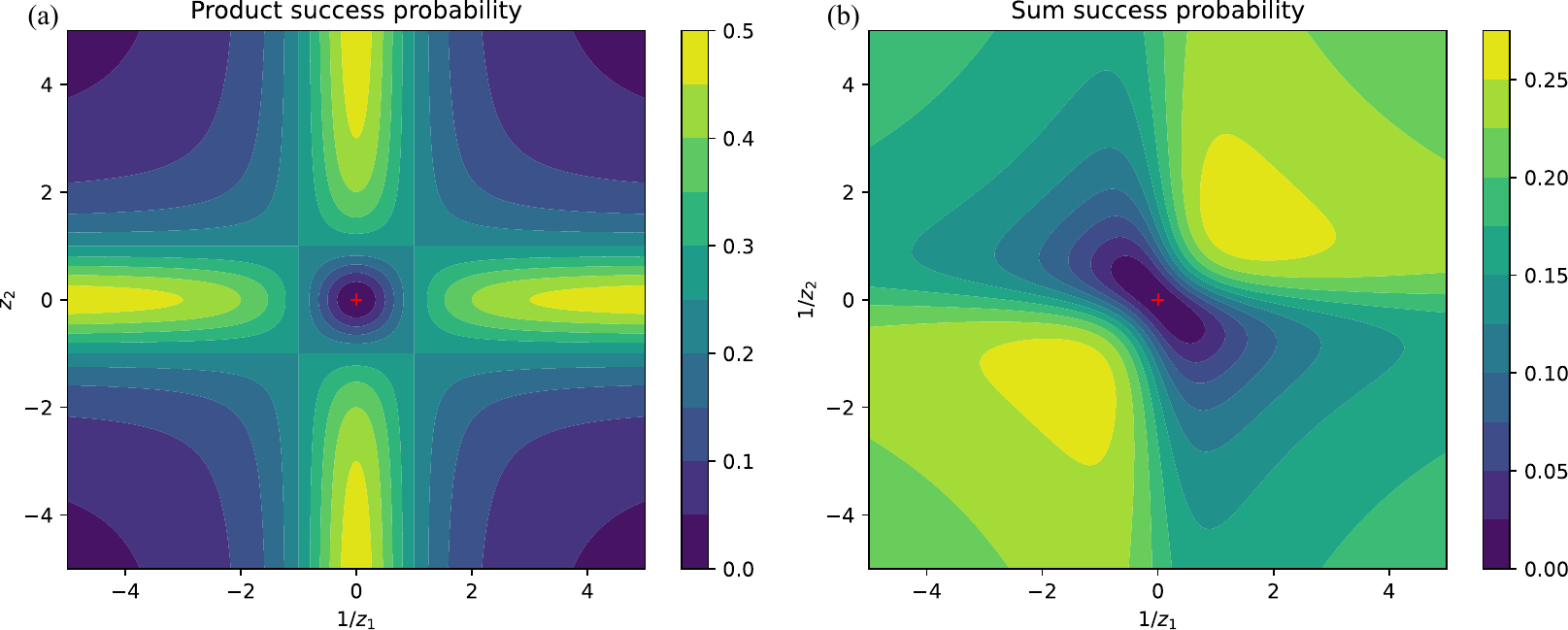}
    \caption{\textbf{Success probability of the operations around the critical points.} a) Success probability of the product operation around the critical point $(\infty,0)$ expressed as a function of $z_1$ and $1/z_2$. b) Success probability of the sum operation around the critical point $(\infty,\infty)$ as a function of $1/z_1$ and $1/z_2$. In both figures, the critical point where the success probability is zero is represented as a red cross.}
    \label{fig:contour}
\end{figure}
In this section, we provide a further study of the success probability of the sum and of the product. In particular, we show that these operations have zero probability of success only for those states in which they are indeterminate such as $(\infty, \infty)$ for the sum and $(0,\infty)$ for the product. Supplementary Fig.~\ref{fig:contour} shows the trends of the success probabilities for the two operations of sum and product, in the region close to these respective critical points. In Supplementary Fig.~\ref{fig:contour}(a) we report the probability of success of the product expressed as a function of $z_1$ and $1/z_2$, considered as real-valued parameters for ease of representation. In this representation, we expect zero probability for the point $(0, 0)$. Analogous analyses are shown in Supplementary Fig.~\ref{fig:contour}(b) for the sum operation.


{
\section{Interferometer designs and their uniqueness} 
\begin{figure}[b!]
    \centering
    \includegraphics{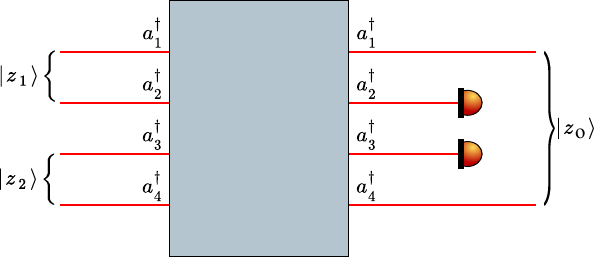}
    \caption{{\textbf{General scheme for two-qubit, four-mode photonic operations.}}}
    \label{fig:general_scheme}
\end{figure}
In the following, we present how we obtained the interferometer designs for the sum and product operations, and show the designs are essentially unique if we use only four modes.
Supplementary Fig.~\ref{fig:general_scheme} shows the general structure for a two-qubit photonic operation using a four-mode interferometer. A photon in the top pair of modes is used to encode the first qubit $\ket{z_1} = \frac{z_1a^\dagger_1+a^\dagger_2}{\sqrt{1+\abs{z_1}^2}}$, and a photon in the bottom pair of modes is used to encode the second qubit $\ket{z_2} = \frac{z_2a^\dagger_3+a^\dagger_4}{\sqrt{1+\abs{z_2}^2}}$.
Without loss of generality, we can assume that the output qubit is encoded in a photon at the first or last output modes ($\ket{z_o} = \frac{z_oa^\dagger_1+a^\dagger_4}{\sqrt{1+\abs{z_o}^2}}$), postselected on outcomes where the second photon is detected in the third mode.
In this context, the resulting (un-normalized) state after the operation is
\begin{equation}
\begin{split}
    \ket{\psi_O} = \{&[(u_{12}u_{34}+u_{32}u_{14})a_1^\dagger+(u_{42}u_{34}+u_{44}u_{32})a_4^\dagger]+
                   z_1[(u_{11}u_{34}+u_{31}u_{14})a_1^\dagger+(u_{41}u_{34}+u_{44}u_{31})a_4^\dagger]+\\
                   +z_2&[(u_{13}u_{32}+u_{12}u_{33})a_1^\dagger+(u_{43}u_{32}+u_{42}u_{33})a_4^\dagger]+
                   z_1z_2[(u_{11}u_{33}+u_{31}u_{13})a_1^\dagger+(u_{41}u_{33}+u_{43}u_{31})a_4^\dagger] \}\ket{0},
\end{split}
\end{equation}
where $u_{ij}$ are the elements of the unitary matrix describing the linear-optical transformation applied to creation operators by the interferometer.
This expression, depending on the operation that one aims to implement, gives us conditions on the interferometer matrix.
\subsection*{Product}
The first operation we want to analyze is the product that results in the unnormalized state $\ket{z_1 z_2} = (z_1 z_2a^\dagger_1+a^\dagger_4) \ket{0}$. To implement this operation, the unitary matrix must satisfy the following conditions:
\begin{gather}
u_{12}u_{34}+u_{32}u_{14} = 0, \label{eq:cond_prod_1}\\
u_{11}u_{34}+u_{31}u_{14} = 0, \label{eq:cond_prod_2}\\
u_{41}u_{34}+u_{44}u_{31} = 0, \label{eq:cond_prod_3}\\
u_{13}u_{32}+u_{12}u_{33} = 0, \label{eq:cond_prod_4}\\
u_{43}u_{32}+u_{42}u_{33} = 0, \label{eq:cond_prod_5}\\
u_{41}u_{33}+u_{43}u_{31} = 0, \label{eq:cond_prod_6}\\
u_{42}u_{34}+u_{44}u_{32} = u_{11}u_{33}+u_{31}u_{13} \neq 0. \label{eq:cond_prod_7}    
\end{gather}
To understand which unitary matrices fulfil these conditions, we start with the expression of Eq.~\eqref{eq:cond_prod_1}. Multiplying it by $u_{31}$ and using Eq.~\eqref{eq:cond_prod_2}, we obtain the following condition:
\begin{equation}
    (u_{31}u_{12}-u_{32}u_{11})u_{34} = 0.
\end{equation}
Then, by multiplying the resulting expression by $u_{33}$ and by using Eq.~\eqref{eq:cond_prod_4}, we obtain:
\begin{equation}
    (u_{31}u_{13}+u_{11}u_{33})u_{32}u_{34} = 0.
\end{equation}
Using Eq.~\eqref{eq:cond_prod_7} we finally obtain the condition:
\begin{equation}
    u_{32} = 0 \; \vee \; u_{34} = 0.
\end{equation}
We start by considering the case where $u_{34} = 0$ (for the other, the procedure is analogous). Taking into account the components that should or should not cancel, we derive the following conditions:
\begin{align}
u_{44} \neq 0 &\quad u_{32} \neq 0, \\
u_{14} = 0 &\quad u_{31} = 0, \\
u_{11} \neq 0 &\quad u_{33} \neq 0, \\
u_{41} &= 0.
\end{align}
Applying these conditions leads to a unitary matrix of the form:
\begin{equation}
U = 
\begin{pmatrix}
\underline u_{11} & u_{12} & u_{13} & 0 \\
u_{21} & u_{22} & u_{23} & u_{24} \\
0 & \underline u_{32} & \underline u_{33} & 0 \\
0 & u_{42} & u_{43} & \underline u_{44} 
\end{pmatrix},
\end{equation}
where the elements need to satisfy the following conditions:
\begin{align}
&u_{13}u_{32}+u_{12}u_{33} = 0, \\
&u_{43}u_{32}+u_{42}u_{33} = 0, \\
&u_{44}u_{32} = u_{11}u_{33},
\end{align}
assuming the underlined terms to be non-zero. Using the orthogonality condition between the first and last columns, we obtain:
\begin{equation}
    u_{21} = 0 \; \vee \; u_{24} = 0.
\end{equation}
If we suppose that the case $u_{21} = 0$ holds (the other case is analogous) then, still by orthogonality, the elements must satisfy also $u_{12} = 0$, $u_{13} = 0$ and $u_{11} = e^{i\varphi}$ (with $\varphi$ arbitrary phase), leading to a matrix of the form
\begin{equation}U = 
\begin{pmatrix}
e^{i\varphi} & 0 & 0 & 0 \\
0 & u_{22} & u_{23} & \underline u_{24} \\
0 & \underline u_{32} & \underline u_{33} & 0 \\
0 & u_{42} & u_{43} & u_{44}
\end{pmatrix}.
\end{equation}
Now we focus on the $3\cross 3$ unitary lower-right submatrix. We use the parametrization presented in Ref.~\cite{Bronzan1988}:
\begin{equation}
\scriptsize
\Tilde{U} = e^{i\phi_6}
\begin{pmatrix}
 - \sin \theta_1 \cos \theta_2 \sin \theta_3 e^{i(\phi_1-\phi_3+\phi_5)}-\sin \theta_2 \cos \theta_3 e^{-i(\phi_2+\phi_4)} & \sin \theta_2 \sin \theta_3 e^{-i(\phi_4+\phi_5)} -\sin \theta_1 \cos \theta_2 \cos \theta_3 e^{i(\phi_1+\phi_2-\phi_3)} & \cos\theta_1 \cos \theta_2 e^{i\phi_1} \\
\cos \theta_1 \sin \theta_3 e^{i\phi_5} & \cos \theta_1 \cos \theta_3 e^{i\phi_2} & \sin \theta_1 e^{i\phi_3} \\
\cos \theta_2 \cos \theta_3 e^{-i(\phi_1 + \phi_2)}-\sin \theta_1 \sin \theta_2 \sin \theta_3 e^{i(-\phi_3+\phi_4+\phi_5)}  & -\cos \theta_2 \sin \theta_3 e^{-i(\phi_1+\phi_5)}-\sin \theta_1 \sin \theta_2 \cos \theta_3 e^{i(\phi_2-\phi_3+\phi_4)} & \cos \theta_1 \sin \theta_2 e^{i\phi_4}
\end{pmatrix}
\end{equation}
with $0\leq \theta_1, \theta_2, \theta_3 \leq \pi/2$ and $0 \leq \phi_1, \dots, \phi_6 \leq 2 \pi$. 
Setting the element $u_{34}$ to be equal to zero leads to $\theta_1 = 0$, simplifying the expression of the matrix to 
\begin{equation}
\Tilde{U} = e^{i\phi_6}
\begin{pmatrix}
-\sin \theta_2 \cos \theta_3 e^{-i(\phi_2+\phi_4)} & \sin \theta_2 \sin \theta_3 e^{-i(\phi_4+\phi_5)} & \cos \theta_2 e^{i\phi_1} \\
\sin \theta_3 e^{i\phi_5} &  \cos \theta_3 e^{i\phi_2} & 0 \\
\cos \theta_2 \cos \theta_3 e^{-i(\phi_1 + \phi_2)}  & -\cos \theta_2 \sin \theta_3 e^{-i(\phi_1+\phi_5)} & \sin \theta_2 e^{i\phi_4}
\end{pmatrix}.
\end{equation}
The condition $u_{43}u_{32}+u_{42}u_{33} = 0$ leads to the expression $\cos \theta_2 (\cos^2 \theta_3 - \sin^2 \theta_3) = 0$, that corresponds to the following condition on the parameters:
\begin{equation}
    \theta_2 = \pi/2 \; \vee \; \theta_3 = \pi/4.
\end{equation}
Moreover, the condition $u_{44}u_{32} = u_{11}u_{33}$ leads to the expression $\cos \theta_3 e^{i(\varphi+\phi_2)} = \sin \theta_3 \sin \theta_2 e^{i(\phi_4+\phi_5+\phi_6)}$. Combining this expression with the previous conditions, we can write that:
\begin{equation}
    \theta_2 = \pi/2 \Rightarrow \theta_3 = \pi/4 \quad \text{or} \quad \theta_3 = \pi/4 \Rightarrow \theta_2 = \pi/2,
\end{equation}
meaning that $\theta_2 = \pi/2$ and $\theta_3 = \pi/4$. 
This gives the final form for the unitary matrix
\begin{equation}U = 
\begin{pmatrix}
e^{i(\phi_1+\phi_2-\phi_3)} & 0 & 0 & 0 \\
0 & - \frac{e^{i \phi_2}}{\sqrt{2}} & \frac{e^{i \phi_3}}{\sqrt{2}} & 0 \\
0 & \frac{e^{-i (\phi_3+\phi_4)}}{\sqrt{2}} & \frac{e^{-i (\phi_2+\phi_4)}}{\sqrt{2}} & 0 \\
0 & 0 & 0 & e^{\phi_1}
\end{pmatrix}
\end{equation}
If we start with the condition $u_{32} = 0$ instead of $u_{34} = 0$ and follow a similar procedure, we arrive at the following expression for unitary matrix
\begin{equation}U = 
\begin{pmatrix}
0 & 0 & e^{i(\phi_1+\phi_2-\phi_3)} & 0 \\
\frac{e^{i \phi_3}}{\sqrt{2}} & 0 & 0 & -\frac{e^{i \phi_2}}{\sqrt{2}} \\
\frac{e^{-i (\phi_2+\phi_4)}}{\sqrt{2}} & 0 & 0 & \frac{e^{-i (\phi_3+\phi_4)}}{\sqrt{2}} \\
0 & e^{\phi_1} & 0 & 0
\end{pmatrix}
\end{equation}
This unitary matrix is equivalent to the previous one up to a swap of column positions, corresponding to the exchange of the first qubit with the second one.
This implies that the geometry of the proposed interferometer is unique.
\subsection*{Addition}
The second operation we need to analyze is the sum operation, which returns the (un-normalized) state $\ket{z_1+z_2} = [(z_1+z_2) a_1^\dagger + a_2^\dagger] \ket{0}$.
To implement this operation, the unitary matrix must satisfy the conditions:
\begin{gather}
u_{12}u_{34}+u_{32}u_{14} = 0, \label{eq:cond_sum_1}\\
u_{41}u_{34}+u_{44}u_{31} = 0, \label{eq:cond_sum_2}\\
u_{43}u_{32}+u_{42}u_{33} = 0, \label{eq:cond_sum_3}\\
u_{11}u_{33}+u_{31}u_{13} =0,  \label{eq:cond_sum_4}\\
u_{41}u_{33}+u_{43}u_{31} = 0, \label{eq:cond_sum_5}\\
u_{42}u_{34}+u_{44}u_{32} = u_{11}u_{34}+u_{31}u_{14} =  u_{13}u_{32}+u_{12}u_{33}\neq 0. \label{eq:cond_sum_6}    
\end{gather}
To understand which unitary matrices fulfil these conditions, we start with the expression Eq.~\eqref{eq:cond_sum_5}. Multiplying it by $u_{32}$ and using Eq.~\eqref{eq:cond_sum_3}, we find the following expression:
\begin{equation}
    (u_{41}u_{32}-u_{42}u_{31})u_{33} = 0.
\end{equation}
We then multiply the resulting expression by $u_{34}$. By using Eq.~\eqref{eq:cond_sum_2} we obtain:
\begin{equation}
    (u_{44}u_{32}+u_{42}u_{34})u_{33}u_{31} = 0.
\end{equation}
Using Eq.~\eqref{eq:cond_sum_5} we then obtain the condition:
\begin{equation}
    u_{33} = 0 \; \vee \; u_{31} = 0.
\end{equation}
Regardless of which of the two conditions one considers to be satisfied, by taking into account which elements must be null or non-zero, we arrive at the following expression for the unitary matrix:
\begin{equation}U = 
\begin{pmatrix}
\underline u_{11} & u_{12} & \underline u_{13} & u_{14} \\
u_{21} & u_{22} & u_{23} & u_{24} \\
0 & \underline u_{32} & 0 & \underline u_{34} \\
0 & u_{42} & 0 & u_{44} 
\end{pmatrix}
\end{equation}
where the elements must satisfy:
\begin{align}
&u_{12}u_{34}+u_{32}u_{14} = 0, \\
&u_{42}u_{34}+u_{44}u_{32}= u_{13}u_{32} = u_{11}u_{34},
\end{align}
assuming the underlined terms to be non-zero. Using the condition $u_{12}u_{34}+u_{32}u_{14} = 0$ together with the orthogonality between the first and second column $u_{12}u_{32}^*+u_{34}^*u_{14}=0$, we obtain
the condition:
\begin{equation}
    u_{14} = 0 \; \vee \; (\abs{u_{32}}^2-\abs{u_{34}}^2) = 0.
\end{equation}
Starting with the condition $(\abs{u_{32}}^2-\abs{u_{34}}^2) = 0$, combined with the normalization condition of the third line of $U$, implies that $\abs{u_{32}}^2 = \abs{u_{34}}^2 = 1/2$. The orthogonality between the third and the fourth row than leads to $\abs{u_{42}}^2 = \abs{u_{44}}^2$, which combined with the normalization condition provides the following constraints $\abs{u_{42}}^2 = \abs{u_{44}}^2 = 1/2$. Using the normalization condition for the fourth column of $U$, and the previous results, we can conclude that $u_{14} = 0$. This allows us to use only the condition $u_{14} = 0$ as the starting point. From $u_{12}u_{34}+u_{32}u_{14} = 0$ we can say that $u_{12} = 0$. From the orthogonality relation between the first and the second column, we obtain that $u_{21}$ or $u_{22}$ must be zero. However, if $u_{21} = 0$, then $u_{13}$ have to be zero, thus contradicting the previous conditions. This implies that $u_{22} = 0$. By the same reasoning, it follows that $u_{24}=0$, leaving with the final matrix
\begin{equation}U = 
\begin{pmatrix}
\underline u_{11} & 0 & \underline u_{13} & 0 \\
\underline u_{12} &  0 & \underline u_{23} & 0 \\
0 & \underline u_{32} & 0 & \underline u_{34} \\
0 & \underline u_{42} & 0 & \underline u_{44}
\end{pmatrix}.
\end{equation}
Combining this expression with the unitary condition, we can parametrize the matrix in the form
\begin{equation}
\label{eq:Uform_supp}
U = 
\begin{pmatrix}
\cos \theta_1 e^{i\phi_1} & 0 & \sin \theta_1 e^{i\phi_2} & 0 \\
\sin \theta_1 e^{-i(\phi_2+\phi_5)} & 0 & -\cos \theta_1 e^{-i(\phi_1+\phi_5)} & 0 \\
0 & \cos \theta_2 e^{i\phi_3} & 0 & \sin \theta_2 e^{i\phi_4} \\
0 & \sin \theta_2 e^{-i(\phi_4+\phi_6)} & 0 & -\cos \theta_2 e^{-i(\phi_3+\phi_6)}
\end{pmatrix}
\end{equation}
with $0< \theta_1, \theta_2< \pi/2$ and $0\leq \phi_1, \dots, \phi_6 \leq 2\pi$.
The constraints $u_{42}u_{34}+u_{44}u_{32}= u_{13}u_{32} = u_{11}u_{34}$ give us the conditions
\begin{equation}
   \phi_1 + \phi_4 = -\phi_6; \quad \phi_2 + \phi_3 = - \phi_6;  \quad \theta_1 = \theta_2; \quad \cos \theta_2 = \sqrt{\frac{5+\sqrt{5}}{10}}; \quad \sin \theta_2 = -\sqrt{\frac{5-\sqrt{5}}{10}}.
\end{equation}
By substituting these conditions in Eq. (\ref{eq:Uform_supp}), we obtain the following expression for the unitary matrix 
\begin{equation}U = \frac{1}{\sqrt{10}}
\begin{pmatrix}
-\sqrt{5+\sqrt{5}} e^{i\phi_1} & 0 & \sqrt{5-\sqrt{5}} e^{i\phi_2} & 0 \\
\sqrt{5-\sqrt{5}} e^{-i(\phi_2+\phi_5)} & 0 & \sqrt{5+\sqrt{5}} e^{-i(\phi_1+\phi_5)} & 0 \\
0 & -\sqrt{5+\sqrt{5}} e^{-i(\phi_2+\phi_6)} & 0 & \sqrt{5-\sqrt{5}} e^{-i(\phi_1+\phi_6)} \\
0 & \sqrt{5-\sqrt{5}} e^{i\phi_1} & 0 & \sqrt{5+\sqrt{5}} e^{i\phi_2}
\end{pmatrix}.
\end{equation}
This implies that the geometry of the proposed interferometer for this operation is unique.}


\section{Experimental apparatus}

The complete experimental layout is reported in Supplementary Fig. \ref{fig:apparato}. One or two pairs of photons are generated in a Beta Barium-Borate crystal. The photons are filtered with a band-pass filter and divided into four different spatial modes using half wave-plates and polarizing beam-splitters. The generated photons are injected in four distinct single-mode fibers, after the polarization and the time-arrival of the photons are compensated via polarization controllers and delay lines so that the photons are made indistinguishable. For the two photon experiments, two input photons are injected into the chip, while the other two input paths are blocked. For the three-photon experiments, three of the four photons are injected into the chip, while the remaining one is used as a trigger. Each output of the chip is connected to a probabilistic number-resolving detector, consisting of six in-fiber beamsplitters and two single-photon avalanche-photodiode detectors for each output of the multimode fiber beam-splitter. All the detectors are connected to a time-to-digital converter for the reconstruction of the output probability.

\begin{figure*}[ht!]
    \centering
    \includegraphics[width=0.99\textwidth]{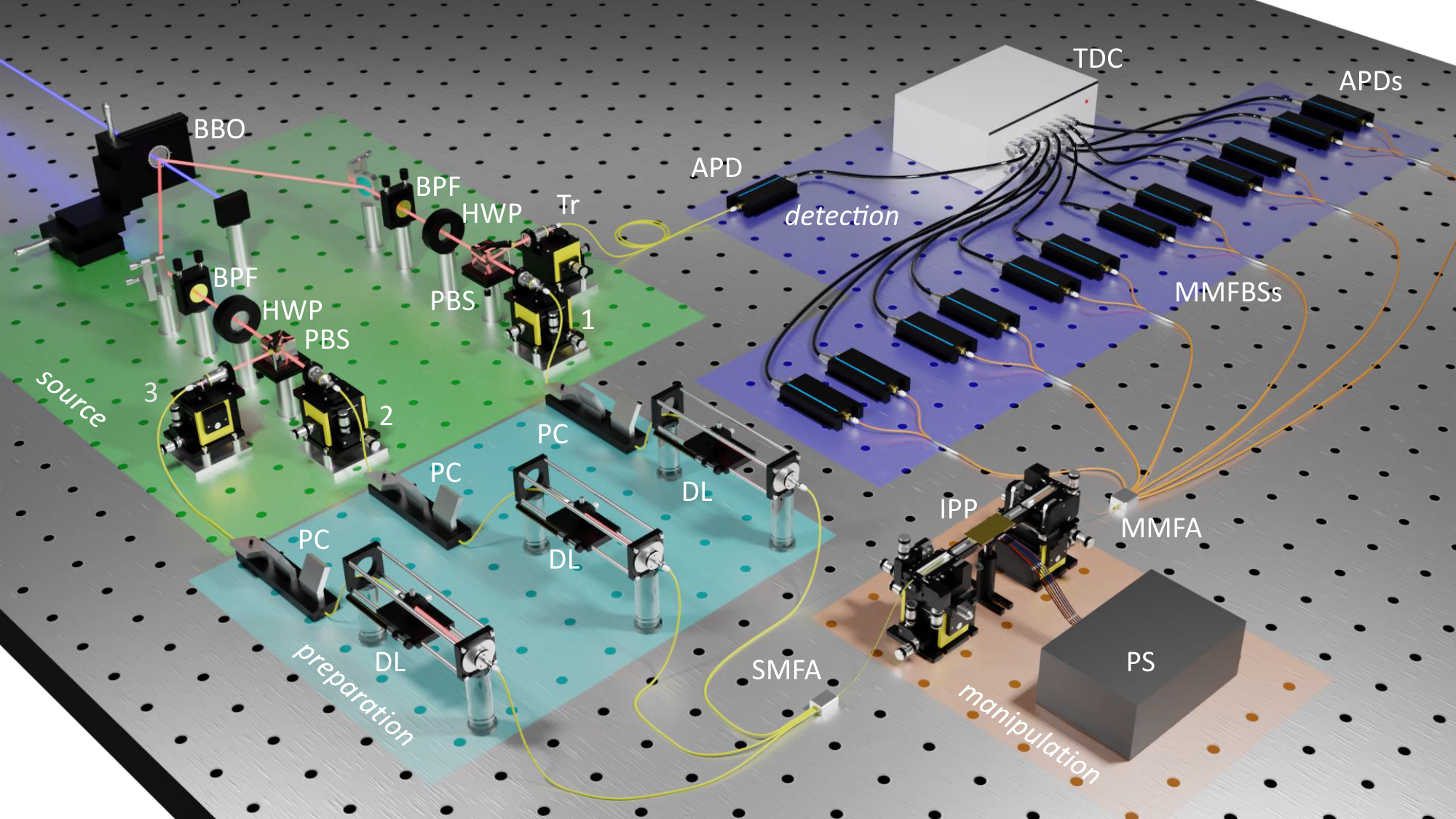}
    \caption{
    \textbf{Scheme of the experimental apparatus.} Complete scheme of the experimental apparatus comprising the photon source, a preparation stage for the input photons, the integrated device implementing the QQBF paradigm, and a detection stage. Legend: BBO - beta-barium borate crystal, BPF - band-pass filter, HWP - half-wave plate,  PBS - polarizing beamsplitter, PC - polarization controller, DL - delay line, PS - power supply, ID- integrated photonic circuit, MMFBS - multi-mode fiber beam-splitter, APD - Single photon avalanche diode detector, TDC - time-to-digital converter.}
    \label{fig:apparato}
\end{figure*}

\section{State preparation and measurement}
In order to perform the experiment it is necessary to prepare the state in the dual-rail encoding. This is performed in the state preparation stage of our device via a programmable beam splitter followed by a phase shifter. In our setting, the programmable beam splitters are implemented through Mach-Zehnder interferometers, having a phase shifter in one of the internal arms (see Supplementary Fig.~\ref{fig:state_preparation_and_projection}~(a)). The implemented matrix $BS(\theta, \phi)$ is:
\begin{equation}
BS(\theta, \phi) = i e^{\frac{\theta}{2}}
    \begin{pmatrix}
    \sin \frac{\theta}{2} e^{i\phi} & \cos \frac{\theta}{2} e^{i\phi} \\
    \cos \frac{\theta}{2} & -\sin \frac{\theta}{2}
    \end{pmatrix}.
\end{equation}
By injecting a single photon in the upper arm of the interferometer, this transformation prepares a general qubit state in the dual-rail scheme:
\begin{equation}
\label{eq:bloch}
    \ket{\theta, \phi} =  \sin\frac{\theta}{2} e^{i\phi} \ket{0} + \cos \frac{\theta}{2} \ket{1},
\end{equation}
that is a generic state in the Bloch sphere. 

\begin{figure*}[t!]
    \centering
    \includegraphics[width=0.85\textwidth]{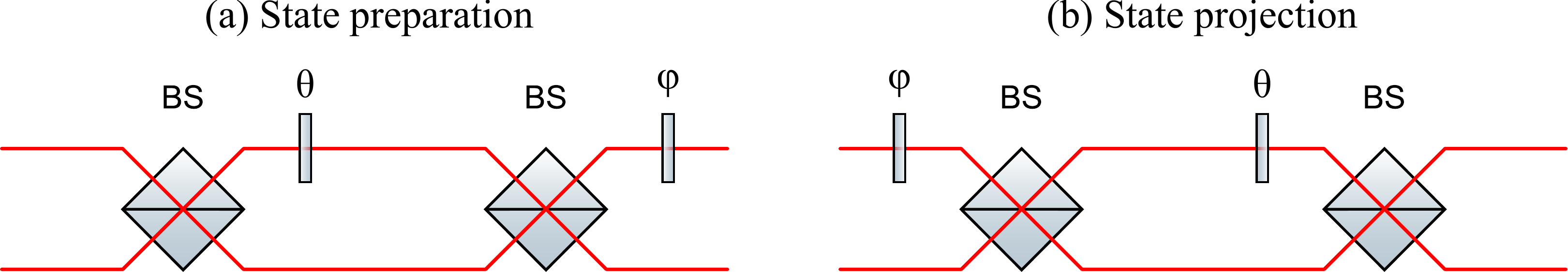}
    \caption{\textbf{Scheme of the state generation and projection steps.} Interferometers for arbitrary state generation (a), and state projection (b). Both schemes are composed by a Mach-Zehnder interferometer with balanced beam splitters with an internal tunable phase $\theta$, with an additional phase shift $\phi$ placed after (state preparation) or before (state projection) the interferometer.}
    \label{fig:state_preparation_and_projection}
\end{figure*}

This scheme and parametrization was employed to generate random value of the complex parameter $z$ in the state parametrization $\vert \mathbf{z} \rangle := \frac{z\ket{0}+\ket{1}}{\sqrt{1+\abs{z}^2}}$, such that the associated states are uniformly distributed on the Bloch sphere (this corresponds to the unique, uniform Haar measure). According to Eq. \eqref{eq:bloch}, the complex parameter $z$ can be directly related to those associated to the Bloch sphere as:
\begin{equation}
    z = \tan \frac{\theta}{2}e^{i\phi} = \sqrt{\frac{1-\cos \theta}{1+\cos \theta}} e^{i\phi}.
\end{equation}
Since the parameters $(\theta, \phi)$ are also the angles of the spherical coordinates, sampling uniformly on the unitary sphere is obtained by sampling uniformly in the rectangle generated by the variable $\phi \in [0,2\pi)$ and $h \in (-1,1]$, where $h = \cos \theta$. Hence, the complex $z$ parameters associated to random states are distributed as:
\begin{equation}
    z = \sqrt{\frac{1-h}{1+h}} e^{i\phi} \qquad (h,\phi) \in (-1,1]\cross[0,2\pi) \; \; \text{uniformly}. 
\end{equation}

Finally, projection on a generic state is obtained by application of a phase shifter followed by a programmable beam splitter [see Supplementary Fig.~\ref{fig:state_preparation_and_projection}~(b)], the associated matrix describing the unitary dynamics is then:
\begin{equation}
BS'(\theta, \phi) = i e^{\frac{\theta}{2}}
    \begin{pmatrix}
    \sin \frac{\theta}{2} e^{i\phi} & \cos \frac{\theta}{2} \\
    \cos \frac{\theta}{2} e^{i\phi} & -\sin \frac{\theta}{2}
    \end{pmatrix}.
\end{equation}
Application of the transformation $BS'(\theta, \phi)$ performed by this interferometer to a generic dual-rail state, followed by a measurement of the output probability on the first mode, is equivalent to projecting the input state onto state
\begin{equation}
    \ket{\theta, -\phi} = \sin\frac{\theta}{2} e^{-i\phi} \ket{0} + \cos \frac{\theta}{2} \ket{1}.
\end{equation}
This allows to obtain direct information on the fidelity of the output states.

\section{Structure and configuration of the photonic processor}
In the main text we reported the schemes (Fig.~4) for the internal configuration of the photonic processor, required to perform the product and addition operations, as well as the concatenation of product followed by a sum. 
In Supplementary Figs.~\ref{fig:chip_structure_inversion}-\ref{fig:chip_structure_sum_sum} we report the internal configuration of the processor to implement and characterize the inversion operation, and other possible concatenations of two operations. Supplementary Fig.~\ref{fig:apparato} provides further details of the physical realization of the photonic device.

\begin{figure*}[ht!]
    \centering
    \includegraphics[scale = 0.20]{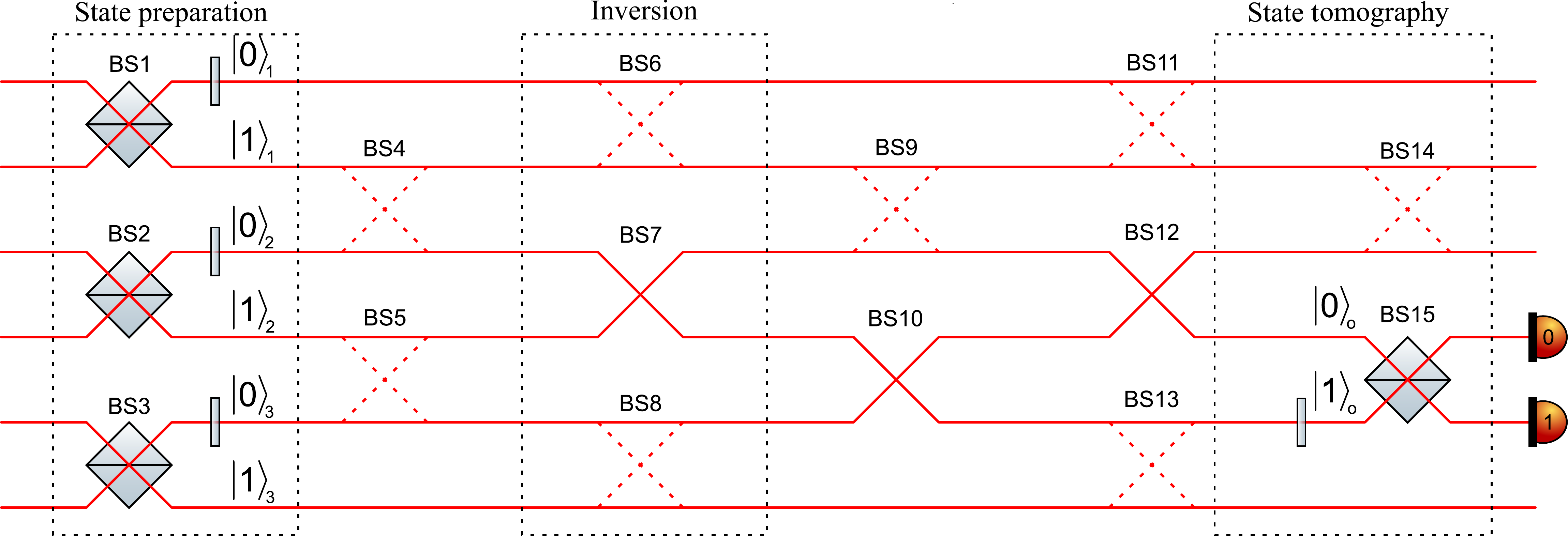}
    \caption{\textbf{Inversion Operation.} Internal configuration of the integrated processor to perform the inversion operation.}
    \label{fig:chip_structure_inversion}
\end{figure*}

\begin{figure*}[ht!]
    \centering
    \includegraphics[scale = 0.20]{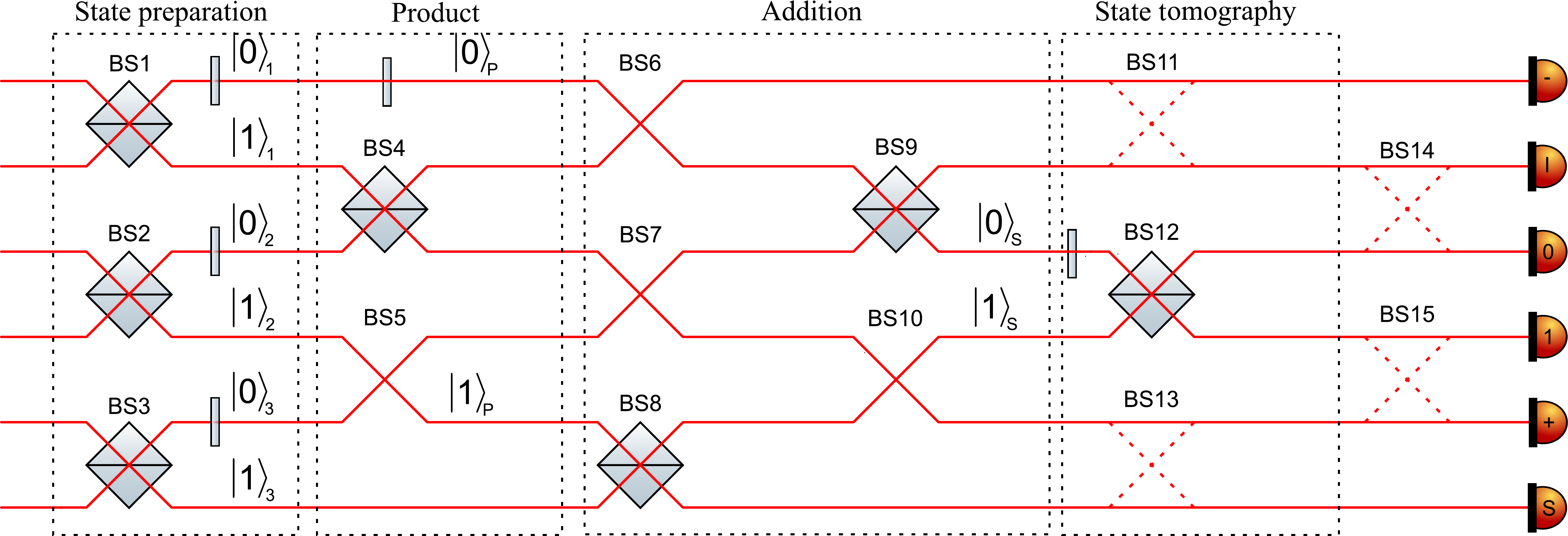}
    \caption{\textbf{Product-Addition Operation.} Internal configuration of the integrated processor to perform the concatenation of a product operation, followed by an addition. Note how, depending on which detectors click, combinations of product or anti-product followed by sum or harmonic mean can be implemented.}
    \label{fig:chip_structure_prod-sum}
\end{figure*}

\begin{figure*}[ht!]
    \centering
    \includegraphics[scale = 0.20]{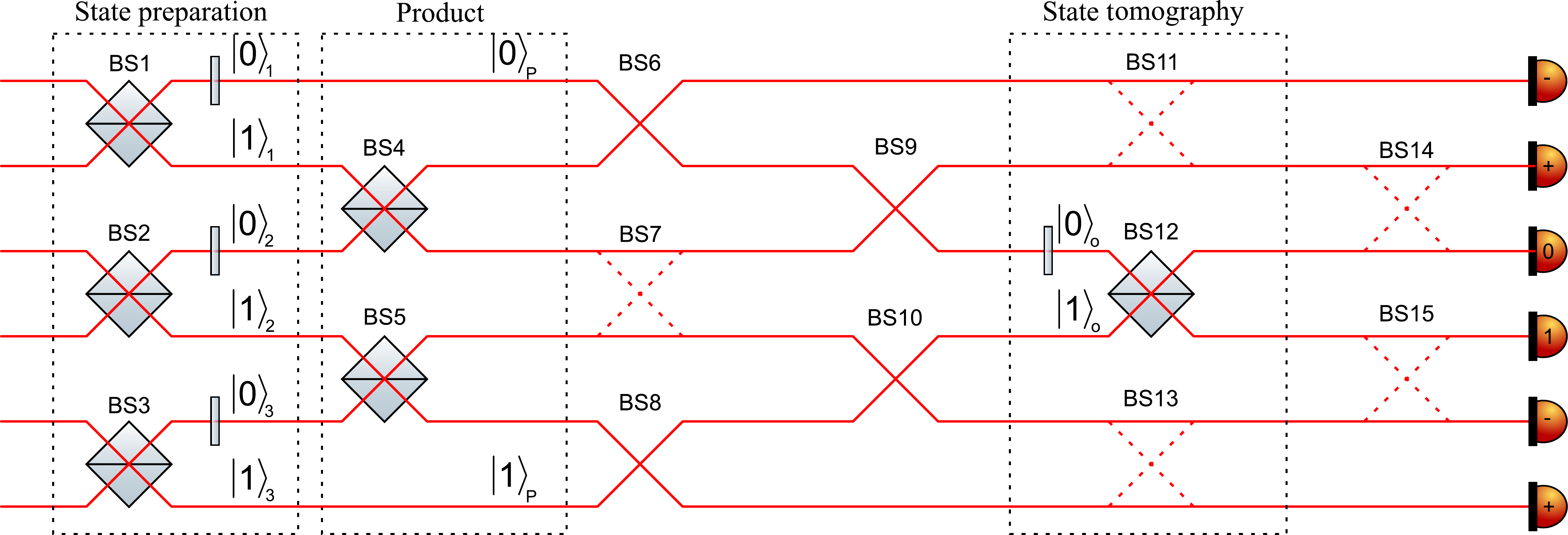}
    \caption{\textbf{Product-Product Operation.} Internal configuration of the integrated processor to perform the concatenation of two product operations.}
    \label{fig:chip_structure_prod_prod}
\end{figure*}

\begin{figure*}[ht!]
    \centering
    \includegraphics[scale = 0.20]{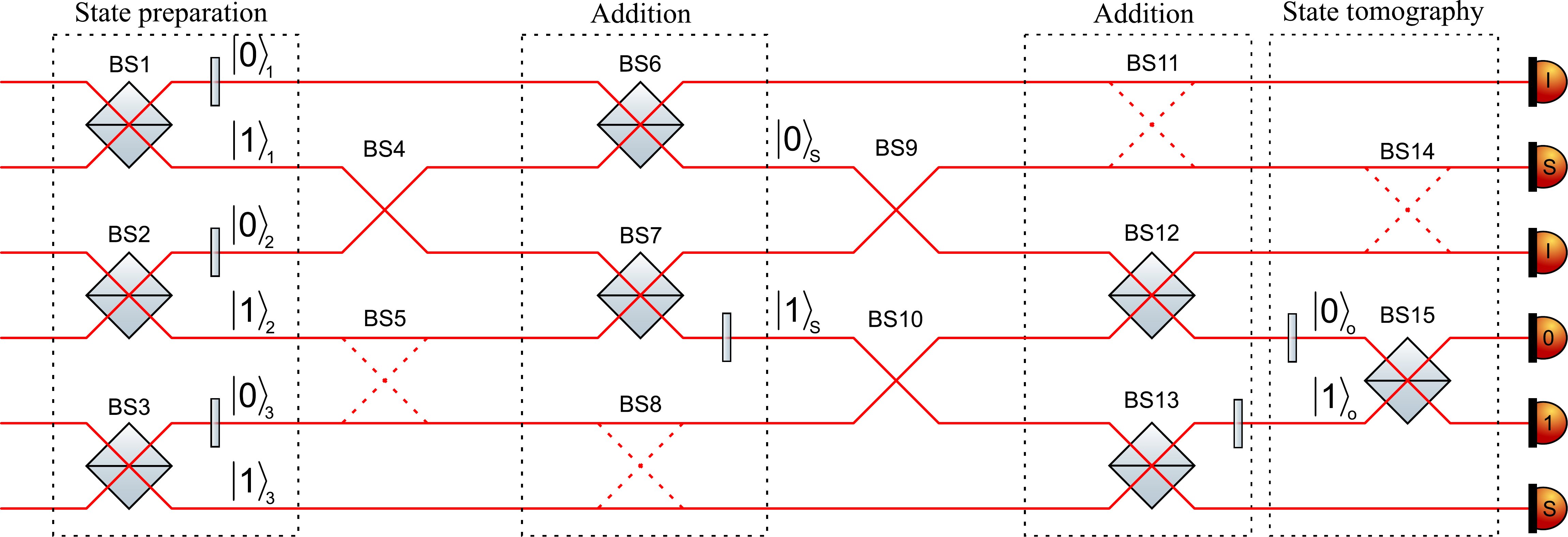}
    \caption{\textbf{Addition-Addition Operation.} Internal configuration of the integrated processor to perform the concatenation of two addition operations.}
    \label{fig:chip_structure_sum_sum}
\end{figure*}

\section{Model with partial photon indistinguishability}

Partial photon indistinguishability between the input photons reduces the fidelity between the output state and the target one. This effect is due to the input source, and is not related to the specific implementation of the QQBF. Here, we discuss a model which considers the effect of limited photon indistinguishability in the predictions for the expected fidelities of the Bernoulli factory implementation. More specifically, we derive an expression for the density matrix of the output state, and the fidelity with respect to the target state, as a function of the input state parameters and of the overlap $C_I$ between the wave functions of the input photon pairs. The indistinguishable case is retrived for $C_I = 1$. We discuss below the case of the building blocks, corresponding to two-photon experiments.
        
\textbf{Product operation.} For the product building block operation, the density matrix of the output state can be written as $\rho_\pm = M_{\pm} \mathcal{N}_{\pm}$, where:
\begin{eqnarray}
    \label{eq:prod_par_1}
    M_{\pm} &=& \begin{pmatrix} 
        \abs{z_1z_2}^2 & \pm z_1z_2C_I \\
        \pm z_1^*z_2^*C_I & 1\\
    \end{pmatrix}, \\
    \mathcal{N}_{\pm} &=&
    \frac{1}{1+\abs{z_1z_2}^2}.
    \label{eq:prod_par_2}
\end{eqnarray}
The fidelity of the output state with respect to the target one is given by:
\begin{equation}
    F_\pm = 1-\frac{2\abs{z_1z_2}^2}{(1+\abs{z_1z_2}^2)^2}(1-C_I).
\end{equation}
We observe that, for the product operation, only the off-diagonal terms of the density matrix are affected by the partial distinguishability of the two input photons.
        
\textbf{Addition operation.} For the addition building block operation, the density matrix of the output state after the addition can be written as $\rho_{S} = M_{S} \mathcal{N}_{S}$, where the matrix elements $(M_S)_{ij}$ of the $2 \times 2$ matrix $M_S$ are:
\begin{eqnarray}
\label{eq:sum_par_1}
(M_S)_{11} &=& RT\abs{z_1+z_2}^2-RT(z_1z_2^*+z_1^*z_2)(1-C_I),\\
(M_S)_{12} &=& \sqrt{RT}(R-T) (z_1+z_2) +\sqrt{RT}(T z_2-R z_1)(1-C_I),\\
(M_S)_{21} &=& \sqrt{RT}(R-T) (z_1^*+z_2^*) +\sqrt{RT}(T z_2^*-R z_1^*)(1-C_I),\\
(M_S)_{22} &=& (R-T)^2+2RT(1-C_I),
\end{eqnarray}
and:
\begin{equation}
\label{eq:sum_par_5}
    \mathcal{N}_{S} = \frac{1}{RT\abs{z_1+z_2}^2+(R-T)^2+RT(2-z_1z_2^*-z_1^*z_2)(1-C_I)}.
\end{equation}
The fidelity of the output state with the target one is given by:
\begin{equation}
    F_s =  1-\frac{2R T\abs{Rz_1+Tz_2}^2}{\biggl[RT\abs{z_1+z_2}^2+(R-T)^2+RT(2-z_1z_2^*-z_1^*z_2)(1-C_I)\biggr]\biggl[RT\abs{z_1+z_2}^2+(R-T)^2\biggr]}(1-C_I).
\end{equation}
        
Conversely, when the harmonic mean is performed, the density matrix of the output state after the addition can be written as $\rho_{I} = M_{I} \mathcal{N}_{I}$, where the matrix elements $(M_I)_{ij}$ of the $2 \times 2$ matrix $M_I$ are:
\begin{eqnarray}
(M_I)_{11} &=& \abs{z_1z_2}^2(R-T)^2+\abs{z_1z_2}^2 2RT(1-C_I),\\
(M_I)_{12} &=& \sqrt{RT}(T-R)(\abs{z_1}^2z_2+z_1\abs{z_2}^2)+\sqrt{RT}(R\abs{z_1}^2z_2-Tz_1\abs{z_2}^2)(1-C_I),\\
(M_I)_{21} &=& \sqrt{RT}(T-R)(\abs{z_1}^2z_2^*+z_1^*\abs{z_2}^2)+\sqrt{RT}(R\abs{z_1}^2z_2^*-Tz_1^*\abs{z_2}^2)(1-C_I),\\
(M_I)_{22} &=& RT \abs{z_1+z_2}^2-RT(z_1z_2^*+z_1^*z_2)(1-C_I),
\end{eqnarray}
and:
\begin{equation}
    \mathcal{N}_{I} = \frac{1}{\abs{z_1z_2}^2(R-T)^2+RT \abs{z_1+z_2}^2+RT(2\abs{z_1z_2}^2-z_1z_2^*-z_1^*z_2)(1-C_I)}.
\end{equation}
The fidelity with the target state reads:
\begin{equation}
F_I = 1-\frac{2RT\abs{R z_1+T z_2}^2\abs{z_1z_2}^2}{\biggl[\abs{z_1z_2}^2(R-T)^2+RT \abs{z_1+z_2}^2+RT(2\abs{z_1 z_2}^2-z_1z_2^*-z_1^*z_2)(1-C_I)\biggr]\biggl[\abs{z_1z_2}^2(R-T)^2+RT\abs{z_1+z_2}^2\biggr]}(1-C_I).
\end{equation}
        
In both cases, it is possible to observe that both diagonal and off-diagonal components of the state are affected by partial photon distinguishability. Hence, limited photonic indistinguishability has a larger impact on the addition operation than it has for the product operation.

\textbf{Assessment of photon indistinguishability in three-photon experiments.} To evaluate the indistinguishability parameter, we have estimated the overlap $C_I$ of the generated photons from Hong-Ou-Mandel interference measurement. In our apparatus, photon pairs were generated via parametric down-conversion in a crystal of beta barium borate (BBO). The overlap must be estimated between photons belonging to the same pair, and for photons belonging to different pairs. The obtained value, after polarization compensation and path synchronization, were respectively $C'_I = 0.99\pm 0.01$ (same pair) and $C''_I = 0.90\pm0.01$ different pairs.

The two-photon experiments performed to demonstrate the building blocks were performed with photons belonging to the same pair. Given the high value of $C'_{I}$ in this case, partial photon indistinguishability does not affect the output fidelity much. For the three-photon experiments, one of the photons belongs to a different pair. For this case, the dominant source of noise is provided by $C''_I$. Hence, in our estimation of $F_D$ in Tab. 1 of the main text, we have neglected the contribution of $C'_I$ and added only the partial distinguishability between photons of different pairs in the theoretical model.

{
\subsection*{Distinguishable particles scenario}
In this subsection we discuss the operation of our schemes when fully distinguishable particles are used as input states.

\textbf{Inversion operation.} The inversion operation does not rely on photon interference, being implemented by flipping the optical modes. This means that the inversion operation is correctly implement even if the input photons are perfectly distinguishable.

\textbf{Product operation.} For the product, the output from the building block can be retrieved starting from Eqs. \eqref{eq:prod_par_1}-\eqref{eq:prod_par_2}, by setting the indistinguishability parameter $C_I = 0$. The resulting density matrix will have zero off-diagonal elements. The diagonal elements of the density matrix, on the other hand, remain unchanged. This means that a measurement in the computational basis will provide an outcome corresponding to the product of the squared moduli of the coefficients $\vert z_1 \vert^2 \vert z_2 \vert^2$.

\textbf{Addition operation.} For the building block that implements addition, the output can be retrieved from Eqs. \eqref{eq:sum_par_1}-\eqref{eq:sum_par_5}, by setting $C_I = 0$. In this case, we first observe that the off-diagonal elements of the density matrix in the computational basis do not reduce to zero. Furthermore, the diagonal elements of the density matrix have the form:
\begin{eqnarray}
(\mathcal{M}_{S})_{11} &=& \frac{1}{\mathcal{N}_S} \frac{RT}{R^2 + T^2} (\vert z_1 \vert^2 + \vert z_2 \vert^2), \\
(\mathcal{M}_{S})_{22} &=& \frac{1}{\mathcal{N}_S},
\end{eqnarray}
where $\mathcal{N}_{S}$ is the normalization factor.
This means that a measurement in the computational basis will provide an outcome corresponding to the sum of the square moduli of the coefficients $\vert z_1 \vert^2 + \vert z_2 \vert^2$ (up to a state independent multiplicative factor).

\textbf{Concatenation of the building blocks.} Let us now discuss the concatenation of the product-addition building blocks when the system is fed with distinguishable particles. We will not discuss concatenation involving the inversion operation, as we have already argued that a deterministic inversion operation is achievable using distinguishable photons.
\begin{itemize}
\item \textit{Product-product.} By concatenating two product operations, which is obtained by a sequence of the corresponding building blocks, with distinguishable input photons, the resulting density matrix is found to have zero-valued off-diagonal elements in the computational basis, and the diagonal elements have the form: 
\begin{eqnarray}
(\mathcal{M}_{PP})_{11} &=& \frac{1}{\mathcal{N}_{PP}} \vert z_1 \vert^2 \vert z_2 \vert^2 \vert z_3 \vert^2, \\
(\mathcal{M}_{PP})_{22} &=& \frac{1}{\mathcal{N}_{PP}},
\end{eqnarray}
where $\mathcal{N}_{PP}$ is the normalization factor.
Hence, a measurement in the computational basis will provide an outcome corresponding to the product of the squared moduli of the three coefficients.
\item \textit{Sum-sum.} By concatenating two sums with distinguishable input particles, the off-diagonal elements of the density matrix in the computational basis do not reduce to zero. Furthermore,  the diagonal elements have the form:
\begin{eqnarray}
(\mathcal{M}_{SS})_{11} &=& \frac{1}{\mathcal{N}_{SS}} \frac{R^2 T^2}{(R^2 + T^2)^2} (\vert z_1 \vert^2 + \vert z_2 \vert^2 + \frac{R^2 + T^2}{RT} \vert z_3 \vert^2, \\
(\mathcal{M}_{SS})_{22} &=& \frac{1}{\mathcal{N}_{SS}},
\end{eqnarray}
where $\mathcal{N}_{SS}$ is the normalization factor.
Hence, the outcome is related to the sum of the squared moduli of the three coefficients (up to state-independent multiplicative factors deriving from the sum building block).
\item \textit{Sum-product.} By concatenating a sum and a product with distinguishable input particles, the off-diagonal elements of the density matrix in the computational basis are found to be zero. Furthermore, the diagonal elements have the form:
\begin{eqnarray}
(\mathcal{M}_{SP})_{11} &=& \frac{1}{\mathcal{N}_{SP}} \frac{R T}{(R^2 + T^2)} (\vert z_1 \vert^2 + \vert z_2 \vert^2) \vert z_3 \vert^2, \\
(\mathcal{M}_{SP})_{22} &=& \frac{1}{\mathcal{N}_{SP}},
\end{eqnarray}
where $\mathcal{N}_{SP}$ is the normalization factor.
Hence, the outcome output is the sum-product concatenation of the squared moduli of the coefficients (up to a state-independent multiplicative factor deriving from the sum building block).
\item \textit{Product-sum.} By concatenating a product and a sum with distinguishable input particles, the off-diagonal elements of the density matrix in the computational basis do not reduce to zero. Furthermore, the diagonal elements have the form:
\begin{eqnarray}
(\mathcal{M}_{PS})_{11} &=& \frac{1}{\mathcal{N}_{PS}} \frac{R T}{(R^2 + T^2)} (\vert z_1 \vert^2 \vert z_2 \vert^2 + \vert z_3 \vert^2), \\
(\mathcal{M}_{PS})_{22} &=& \frac{1}{\mathcal{N}_{PS}},
\end{eqnarray}
where $\mathcal{N}_{PS}$ is the normalization factor.
Hence, the output is the result of the product-sum concatenation of the squared moduli of the coefficients (up to a state-independent multiplicative factor).
\end{itemize}
In summary, the distinguishable particle scenario provides, after the measurement in a computational basis, an output bit which is compatible with the desired transformation applied not on the $z_i$ input parameters themselves, but rather on the squared moduli of these coefficients. The multiplicative factors $R T/(R^2 + T^2)$, arising from the sum building block, are irrelevant, since the are state-independent and can be then compensated by adding additional product modules with an input state with coefficient $z_i = \sqrt{(R^2+T^2)/(RT)}$.
}

\section{Data analysis}
In the reported experiment, the mean fidelity of the output states with respect to the target one after the different QQBF operations is employed to assess the quality of the implementation. In particular, the mean is performed over a set of pairs of input states sampled uniformly from the Bloch sphere. In particular, for each pair we program the first layer of the chip to implement the desired states and the last layer to perform a projective measure on the expected state to estimate the fidelity.

In the two photons case, and with reference to the circuit layouts, we program the chip in the prescribed structure according to the required QQBF operation and we collect the coincidences at the output. The state fidelity between the measured states and the target one is calculated as:
\begin{equation}
    F_M = \frac{C_{0y}}{C_{0y}+C_{1y}},
\end{equation}
where $C_{0y}$ and $C_{1y}$ are the coincidences respectively between the output $0$ or $1$ and the output required to post-select on the correct operation, that is $y=+$ for the product and $y=S$ for the sum (see Fig.~4 of the main text). The fidelity $F_M$ is evaluated directly from the raw data. 

The value of $F_C$ is then obtained by subtracting the coincidences due to spurious events present in the apparatus associated with dark counts and accidental coincidences. These events are estimated from the singles counts as: 
\begin{equation}
    \tilde{C}_{xy} = \frac{N_x N_y}{f T},
\end{equation}
where $N_x$ is the singles count measured in channel $x$,  $N_y$ is the singles counts measured on the detector corresponding to the post-selection output ($y=+$ or $y=S$ depending on the QQBF building block operation), $f$ is the repetition rate of the pulsed laser used and $T$ is the exposure time. The fidelity $F_C$ is obtained as: 
\begin{equation}
    F_C = \frac{C_{0y}-\tilde{C}_{0y}}{C_{0y}-\tilde{C}_{0y}+C_{1y}-\tilde{C}_{1y}}.
\end{equation}

An analogous data analysis is also performed for the three-photon experiments.

{The theoretical fidelity $F_D$ is calculated from the models presented in the previous section taking into account only the partial distinguishability between the photons as a source of noise.}

{All the errors are estimated from the propagation of the Poisson statistics proper of single-photon counts.}


%